# The oscillation properties of the Blue Large Amplitude Pulsators (BLAPs): relative change rate of periods, excitations, and period relations

Tao Wu[1,2,3,4,5★] and Yan Li[1,2,3,4,5★]

[1] *Yunnan Observatories, Chinese Academy of Sciences, 396 Yangfangwang, Guandu District, Kunming 650216, P. R. China*
[2] *Key Laboratory for the Structure and Evolution of Celestial Objects, Chinese Academy of Sciences, 396 Yangfangwang, Guandu District, Kunming 650216, P. R. China*
[3] *Center for Astronomical Mega-Science, Chinese Academy of Sciences, 20A Datun Road, Chaoyang District, Beijing 100012, P. R. China*
[4] *University of Chinese Academy of Sciences, Beijing 100049, P. R. China*
[5] *International Centre of Supernovae, Yunnan Key Laboratory, Kunming 650216, P. R. China*

## ABSTRACT

**B**lue **L**arge **A**mplitude **P**ulsators (BLAPs) are a type of variable star that has been identified relatively recently. They are characterized by their large amplitude, high gravity, and long periods (2–75 min). Some BLAPs exhibit a rich helium abundance on their surfaces, while some of the others show rich hydrogen atmospheres. Up to date, both the mode identification and formation pathways of BLAPs remain subjects of ongoing debate. In this work, we gave an non-adiabatic pulsation analyses for low-order radial and non-radial dipole modes on a theoretical model grid to explore the characteristics of BLAPs. The effects of element diffusion and radiative levitation are not taken into account our theoretical models. This is because these processes could lead to discrepancies in surface element abundance of the sdB or BLAP stars between theoretical models and observations. Based on the theoretical model grid, our analyses revealed that the radial fundamental modes align well with various observed properties of BLAPs, including the pulsation instability, the position of the excitable models on the HR diagram, surface element compositions, period ranges, and period changes. Our theoretical models also predicted a period–gravity ($\log g$) relation with a narrow distribution, while the observations exhibit a broader range. Furthermore, our findings indicate that the excitation of pulsation in BLAPs is primarily influenced by metal abundance. The thickness of the outer rich hydrogen envelope and the ratio of helium to hydrogen on the surface almost do not affect the pulsation instability.

**Key words:** asteroseismology – instabilities – stars: evolution – stars: interiors – stars: oscillations – stars: variables: general.

## 1 INTRODUCTION

Asteroseismology plays a crucial role in modern astrophysics. It provides a unique insight into the interiors of stars through the study of oscillation waves. It can be utilized to investigate stellar internal structures and evolutionary states. Therefore, we can test physical processes and calibrate fundamental parameters in stars with asteroseismology, for instance, the convective mixing length, convective overshooting, opacity, nuclear reaction rate, and so on. Ultimately, this approach aims to examine and improve the theory of stellar evolution and structure (refer to the asteroseismology book and the recent reviews, e.g. C. Aerts, J. Christensen-Dalsgaard & D. W. Kurtz 2010; 2013; S. Hekker & J. Christensen-Dalsgaard 2017; C. Aerts, S. Mathis & T. M. Rogers 2019; S. Basu & S. Hekker 2020; C. Aerts 2021; D. W. Kurtz 2022; C. Aerts & A. Tkachenko 2024).

Moreover, asteroseismology offers a method for determining fundamental stellar parameters with unexpected precision, including stellar mass, radius, gravity, distance, and age. These high-precision measurements enhance our understanding of stellar properties and can be applied to various research fields, from exoplanet and cluster to galaxy and cosmology.

**B**lue **L**arge **A**mplitude **P**ulsators (hereafter, BLAPs) are a type of variable star that has been identified relatively recently. They are characterized by large amplitude (on the order of tenths of a magnitude in $V$ band) and long periods (from a few minutes to more than one hour). This pulsation type was first observed and identified by P. Pietrukowicz et al. (2017) during the **O**ptical **G**ravitational **L**ensing **E**xperiment (OGLE) sky survey (for more information, refer to A. Udalski et al. 1992, and related works). The light curves of BLAPs exhibit a sawtooth shape similar to the pulsation patterns of $\delta$ Scuti stars, classical Cepheids, and RR Lyrae stars pulsating in the fundamental mode (refer to fig. 1 of P. Pietrukowicz et al. 2017). Initially, fourteen BLAPs were discovered and catalogued in the OGLE survey over a decade of observations (P. Pietrukowicz et al. 2017). These stars were designated as OGLE-BLAP-001 to OGLE-BLAP-014. The relative period change rate ($\dot{P}/P$) for eleven of them was determined by P. Pietrukowicz et al. (2017) and for the other three (OGLE-BLAP-002, -004, and -011) was supplemented by T. Wu & Y. Li (detailed refers to section 2 of 2018). For all fourteen OGLE-BLAPs, $\dot{P}/P$ is between $-20$ and $+8 \times 10^{-7}$ yr$^{-1}$. Positive values indicate an increasing pulsating period in expanding stars, while negative values suggest a decreasing period in contracting stars. This behavior suggests that BLAPs are in a non-monotonic evolutionary phase.

★ E-mail: wutao@ynao.ac.cn (TW), ly@ynao.ac.cn (YL)

In the work of P. Pietrukowicz et al. (2017), spectroscopic observations of four BLAPs (OGLE-BLAP-001, -009, -011, and -014) indicate that BLAPs are located on the HR diagram at approximately $\log T_{\rm eff}/{\rm K} \approx 4.4$–$4.5$ and $\log g \approx 4.0$–$4.7$. These stars have hydrogen-poor envelopes, with a ratio of helium to hydrogen $\log(N_{\rm He}/N_{\rm H}) \approx -0.5$ at the stellar surface. This suggests that these BLAPs have richer helium than hydrogen. More details refer to P. Pietrukowicz et al. (2017).

Since the discovery of BLAPs by P. Pietrukowicz et al. (2017) and T. Kupfer et al. (2019) has identified a sub-type known as 'high-gravity-BLAP' (hereafter, hg-BLAP), characterized by their higher gravity ($\log g \approx 5.3$–$5.7$) and slightly higher effective temperatures ($\log T_{\rm eff}/{\rm K} \geq 4.5$) compared to the normal BLAPs (OGLE-BLAPs). These hg-BLAPs also exhibit shorter periods ($P \approx 3$–$8$ min) and poorer helium abundance on their surface ($\log(N_{\rm He}/N_{\rm H}) \sim -2$). Subsequent studies by researchers such as P. R. McWhirter & M. C. Lam (2022), A. Pigulski, K. Kotysz & P. A. Kołaczek-Szymański (2022), J. Lin et al. (2022), G. Ramsay et al. (2022), J. Borowicz et al. (2023a, b), S.-W. Chang et al. (2024), and P. Pietrukowicz et al. (2024) have expanded the BLAPs and hg-BLAPs sample size to 131, with periods ranging from 2–75 min, as detailed in Table 1.

In the OGLE fourth survey phase (OGLE-IV), from the observations of Galactic fields, J. Borowicz et al. (2023a) identified 25 BLAPs (e.g. OGLE-BLAP-001 and OGLE-BLAP-038 to OGLE-BLAP-061). 20 of them are newly discovered. In the outer Galactic bulge, J. Borowicz et al. (2023b) found 33 new BLAPs (OGLE-BLAP-062 to OGLE-BLAP-094). Among them, OGLE-BLAP-086 and -093 are also called OW-BLAP-4 (G. Ramsay et al. 2022) and SMSS-BLAP-1 (S.-W. Chang et al. 2024), respectively. Moreover, within the Galactic Bulge region, P. Pietrukowicz et al. (2024) identified 23 BLAPs. They also reported various measurements, including the largest period change $+45.80 \pm 0.65 \times 10^{-7}$ yr$^{-1}$ in OGLE-BLAP-030, and the spectroscopic analyses on OGLE-BLAP-001 and 14 other new BLAPs using the MagE spectrograph to derive atmospheric parameters ($T_{\rm eff}$, $\log g$, and $\log(N_{\rm He}/N_{\rm H})$) as well as the radial velocities.

Among these targets, only three objects (HD 133729, TMTS-BLAP-1, and OGLE-BLAP-007) are confirmed to be in binary systems. HD 133729 accompanies a main-sequence B-type star (more details refers to A. Pigulski et al. 2022) and TMTS-BLAP-1 with an unseen object (more information refers to J. Lin et al. 2022), respectively. For OGLE-BLAP-007, S.-L. Kim et al. (2025) reports that it has two wide-orbit companions. Out of all samples, 35 targets have determined the relative period change rates ($\dot{P}/P$). Their values vary from $-19.23 \times 10^{-7}$ yr$^{-1}$ to $+45.80 \times 10^{-7}$ yr$^{-1}$. 30 of them have completed spectroscopic observations. Additionally, for the ratio between helium and hydrogen $\log(N_{\rm He}/N_{\rm H})$ at stellar surface, 29 targets are measured. Their values vary from $-2.8$ to $-0.41$. All observed parameters are listed in Table 1. To date, there have been 103 confirmed BLAPs (including four hg-BLAPs), seven high-confidence candidates (including four hg-BLAP candidates), and nine candidates (including two hg-BLAP candidates).

Besides the abovementioned confirmed targets and candidates, there are an additional 12 potential candidates found by T. Kupfer et al. (2021). They are designated as ZTF-sdBV1 through ZTF-sdBV12 (refer to Table 1). These candidates exhibit single-period pulsations with amplitudes in tens of mmag in the $r$ band. Their pulsating periods span from approximately 6 to 17 min, which effectively bridge the period gap between normal BLAPs (i.e. OGLE-BLAPs-01 – -14) and hg-BLAPs. Their light curve shapes closely resemble ones of the identified BLAPs. Comprehensive spectroscopic observations are imperative to ascertain whether these candidates are indeed BLAPs or single-pulsating mode sdB stars.

At present, the identification of pulsating modes (i.e. radial or non-radial, fundamental or high-order modes) and the determination of the evolutionary states of the BLAPs remain subjects of intense debate. This controversy is evident in various studies (refers to, for instance, P. Pietrukowicz et al. 2017; C. M. Byrne & C. S. Jeffery 2018; A. D. Romero et al. 2018; T. Wu & Y. Li 2018). Therefore, the evolutionary pathways (or formation channel) of BLAPs are currently also a topic of debate (for further discussions, see P. Pietrukowicz et al. 2017; C. M. Byrne & C. S. Jeffery 2018, 2020; X.-C. Meng et al. 2020; C. M. Byrne, E. R. Stanway & J. J. Eldridge 2021; H. Xiong et al. 2022).

Based on the assumption of that the pulsations of BLAPs are the radial fundamental mode, P. Pietrukowicz et al. (2017) suggested that BLAPs may be either low-mass ($\sim 0.3$ M$_\odot$) shell-hydrogen-burning stars or high-mass cases ($\sim 1.0$ M$_\odot$) undergoing core helium burning. Furthermore, they discussed the instability of pulsations and suggested that both the radial fundamental mode and the first overtone are stability (i.e. growth rate $\eta < 0$, indicating that pulsations cannot be excited). They concluded that the first overtone is even more stable (i.e. $\eta_{1o} < \eta_{\rm F} < 0$, where the subscript '1o' notes 1$^{\rm th}$ overtone, and 'F' notes fundamental mode) P. Pietrukowicz et al. (2017). Similarly, T. Wu & Y. Li (2018) reported that these OGLE-BLAPs are in the core-helium-burning evolutionary phase with masses around $\sim 1.0$ M$_\odot$. Their theoretical models successfully replicate all properties of observed BLAPs at that time, including period $P$, period change rate $\dot{P}/P$, effective temperature $T_{\rm eff}$, surface gravity $\log g$, and helium-to-hydrogen ratio $\log(N_{\rm He}/N_{\rm H})$ at stellar surface. Notably, these models precisely capture the period evolution (both increasing and decreasing) (for more details, see T. Wu & Y. Li 2018, e.g. figs 3–5), as illustrated in Fig. 1. Unfortunately, their work did not discuss pulsation instability.

Fig. 1 shows that the theoretical models used by T. Wu & Y. Li (2018) encompass the pre-core-helium-burning, core-helium-burning, and shell-helium-burning phases. But, in the previous work, they were not distinguished. These models are simply divided into early core-helium-burning and evolved core-helium-burning stages (see fig. 5 of T. Wu & Y. Li 2018, for more details). Fig. 1 indicates that these models cannot account for these targets (including confirmed BLAPs and candidates) with low pulsating period and high gravity. These targets were all discovered later. To address this, models with lower helium core mass and thinner envelopes might be imperative. In the following analysis, we will take into account this factor in models.

The rate of relative period change $\dot{P}/P = -11.5 \pm 0.6 \times 10^{-7}$ yr$^{-1}$ of HD 133729 (A. Pigulski et al. 2022) indicates that it has a faster evolutionary speed compared to most normal OGLE-BLAPs. A. Pigulski et al. (2022) proposed that it shares similarities with the OGLE-BLAPs. In fact, the exact evolutionary state of HD 133729 remains incompletely understood due to the absence of a comprehensive theoretical model analysis in pulsation studies. Referring to the theoretical models by T. Wu & Y. Li (2018), Fig. 1 indicates that HD 133729 is likely in the early core-helium-burning or evolved shell-helium-burning phases. Unfortunately, we cannot even completely exclude the possibility of HD 133729 being in the core-helium-burning phase.

Detailed analysis by J. Lin et al. (2022) suggested that TMTS-BLAP-1 is in a rapidly evolving state of shell-helium burning, with an estimated mass of $\sim 0.7$ M$_\odot$. This is because that TMTS-BLAP-1 has

**Table 1.** Observational Parameters of identified BLAPs.

| ID | $P$ (min) | $\dot{P}/P$ ($10^{-7}$yr$^{-1}$) | $T_{\rm eff}$ (K) | $\log g$ (c.g.s) | $\log(N_{\rm He}/N_{\rm H})$ | Ref. |
|---|---|---|---|---|---|---|
| OGLE-BLAP-001√ | 28.26(28.2549636) | $2.90 \pm 3.70$ | $30800 \pm 500$ | $4.61 \pm 0.07$ | $-0.55 \pm 0.05$ | [1]([9]) |
| OGLE-BLAP-002√ | 23.29 | $-19.23 \pm 8.05$ | – | – | – | [1],[2]♠ |
| OGLE-BLAP-003√ | 28.46 | $0.82 \pm 0.32$ | – | – | – | [1] |
| OGLE-BLAP-004√ | 22.36 | $-5.03 \pm 1.57$ | – | – | – | [1],[2]♠ |
| OGLE-BLAP-005√ | 27.25 | $0.63 \pm 0.26$ | – | – | – | [1] |
| OGLE-BLAP-006√ | 38.02 | $-2.85 \pm 0.31$ | – | – | – | [1] |
| OGLE-BLAP-007√ | 35.18 | $-2.40 \pm 0.51$ | – | – | – | [1] |
| OGLE-BLAP-008√ | 34.48 | $2.11 \pm 0.27$ | – | – | – | [1] |
| OGLE-BLAP-009√ | 31.94 | $1.63 \pm 0.08$ | $31800 \pm 1400$ | $4.40 \pm 0.18$ | $-0.41 \pm 0.13$ | [1] |
| OGLE-BLAP-010√ | 32.13 | $0.44 \pm 0.21$ | $29800 \pm 300$ | $4.57 \pm 0.04$ | $-0.58 \pm 0.03$ | [1],[10]◇ |
| OGLE-BLAP-011√ | 34.87 | $6.77 \pm 8.87$ | $26200 \pm 2900$ | $4.20 \pm 0.20$ | $-0.45 \pm 0.11$ | [1],[2]♠ |
| OGLE-BLAP-012√ | 30.90 | $0.03 \pm 0.15$ | – | – | – | [1] |
| OGLE-BLAP-013√ | 39.33 | $7.65 \pm 0.67$ | – | – | – | [1] |
| OGLE-BLAP-014√ | 33.62 | $4.82 \pm 0.39$ | $30900 \pm 2100$ | $4.42 \pm 0.26$ | $-0.54 \pm 0.16$ | [1] |
| high-gravity-BLAP-1√ | 3.34 | – | $34000 \pm 500$ | $5.70 \pm 0.05$ | $-2.1 \pm 0.2$ | [3] |
| high-gravity-BLAP-2√ | 6.05 | – | $31400 \pm 600$ | $5.41 \pm 0.06$ | $-2.2 \pm 0.3$ | [3] |
| high-gravity-BLAP-3√ | 7.31 | – | $31600 \pm 600$ | $5.33 \pm 0.05$ | $-2.0 \pm 0.2$ | [3] |
| high-gravity-BLAP-4√ | 7.92 | – | $31700 \pm 500$ | $5.31 \pm 0.05$ | $-2.4 \pm 0.4$ | [3] |
| ZGP-BLAP-02 | 48.258 | – | – | – | – | [4] |
| ZGP-BLAP-03 | 53.705 | – | – | – | – | [4] |
| ZGP-BLAP-04 | 46.681 | – | – | – | – | [4] |
| ZGP-BLAP-05♡(OGLE-BLAP-057√) | 54.000(54.6530265) | – | – | – | – | [4]([9]) |
| ZGP-BLAP-06♡(OGLE-BLAP-058√) | 35.839(35.8390155) | – | – | – | – | [4]([9]) |
| ZGP-BLAP-07♡(OGLE-BLAP-060√) | 44.627(44.6277164) | – | – | – | – | [4]([9]) |
| ZGP-BLAP-08♡(OGLE-BLAP-061√) | 35.137(35.1375078) | – | – | – | – | [4]([9]) |
| ZGP-BLAP-09√ | 23.264 | $-1.2 \pm 1.6$ | 35000⊤ | $5.0 \pm 0.25$ | $-2.6 \sim -1.1$§ | [4],[6]♣ |
| ZGP-BLAP-10♡ | 55.180 | – | – | – | – | [4] |
| ZGP-BLAP-11♡ | 27.951 | – | – | – | – | [4] |
| ZGP-BLAP-12 | 51.619 | – | – | – | – | [4] |
| ZGP-BLAP-13 | 21.578 | – | – | – | – | [4] |
| ZGP-BLAP-14♡ | 17.016 | – | – | – | – | [4] |
| ZGP-BLAP-15 | 51.073 | – | – | – | – | [4] |
| ZGP-BLAP-16 | 37.330 | – | – | – | – | [4] |
| ZGP-HGBLAP-01 | 2.428 | – | – | – | – | [4] |
| ZGP-HGBLAP-02♡ | 6.071 | – | – | – | – | [4] |
| ZGP-HGBLAP-03♡ | 4.203 | – | – | – | – | [4] |
| ZGP-HGBLAP-04♡ | 5.950 | – | – | – | – | [4] |
| ZGP-HGBLAP-05 | 8.240 | – | – | – | – | [4] |
| ZGP-HGBLAP-06♡ | 6.260 | – | – | – | – | [4] |
| HD 133729‡,√ | 32.37 | $-11.5 \pm 0.6$ | 29000 | 4.50 | – | [5] |
| TMTS-BLAP-1‡,√(ZGP-BLAP-01√) | 18.90(18.933) | $22.3 \pm 0.9$ | $28810 \pm 2970$ | $4.69 \pm 0.07$ | $-0.61 \pm 0.10$ | [6]([4]) |
| OW-BLAP-1√ | 10.8 | – | $30600 \pm 2500$ | $4.67 \pm 0.25$ | $-2.1 \pm 0.2$ | [7] |
| OW-BLAP-2√ | 23.0 | – | $27300 \pm 1500$ | $4.83 \pm 0.20$ | $-0.7 \pm 0.1$ | [7] |
| OW-BLAP-3√ | 28.9 | – | $29900 \pm 3500$ | $4.16 \pm 0.40$ | $-0.8 \pm 0.3$ | [7] |
| OW-BLAP-4√(OGLE-BLAP-086√) | 32.0(32.9633868) | – | $27300 \pm 2000$ | $4.20 \pm 0.20$ | $-0.8 \pm 0.2$ | [7]([12]) |
| OGLE-BLAP-038√ | 11.2601526 | – | – | – | – | [9] |
| OGLE-BLAP-039√ | 31.9651131 | – | – | – | – | [9] |
| OGLE-BLAP-040√ | 31.4607770 | – | – | – | – | [9] |
| OGLE-BLAP-041√ | 53.5405245 | – | – | – | – | [9] |
| OGLE-BLAP-042√ | 62.0505860 | – | $28300 \pm 1000$ | $4.19 \pm 0.14$ | $-0.52 \pm 0.12$ | [9],[10]◇ |
| OGLE-BLAP-043√ | 19.0642504 | – | – | – | – | [9] |
| OGLE-BLAP-044√ | 8.4216381 | – | $32700 \pm 200$ | $5.28 \pm 0.03$ | $-2.80 \pm 0.09$ | [9],[10]◇ |
| OGLE-BLAP-045√ | 41.3305960 | – | – | – | – | [9] |
| OGLE-BLAP-046√ | 39.4729321 | – | – | – | – | [9] |
| OGLE-BLAP-047√ | 31.2450830 | – | – | – | – | [9] |
| OGLE-BLAP-048√ | 28.8555579 | – | – | – | – | [9] |
| OGLE-BLAP-049√ | 16.4015624 | – | $29300 \pm 400$ | $4.91 \pm 0.06$ | $-0.67 \pm 0.04$ | [9],[10]◇ |
| OGLE-BLAP-050√ | 22.9352027 | – | – | – | – | [9] |
| OGLE-BLAP-051√ | 45.4160893 | – | – | – | – | [9] |
| OGLE-BLAP-052√ | 34.9622342 | – | – | – | – | [9] |

**Table 1** – *continued*

| ID | $P$ (min) | $\dot{P}/P$ ($10^{-7}$yr$^{-1}$) | $T_{\rm eff}$ (K) | $\log g$ (c.g.s) | $\log(N_{\rm He}/N_{\rm H})$ | Ref. |
|---|---|---|---|---|---|---|
| OGLE-BLAP-053√ | 25.2601420 | – | – | – | – | [9] |
| OGLE-BLAP-054√ | 11.3324455 | – | – | – | – | [9] |
| OGLE-BLAP-055√ | 35.3945271 | – | – | – | – | [9] |
| OGLE-BLAP-056√ | 11.6725445 | – | – | – | – | [9] |
| OGLE-BLAP-059√ | 33.5823049 | – | – | – | – | [9] |
| OGLE-BLAP-015√ | 43.073 | $1.91 \pm 0.99$ | – | – | – | [10] |
| OGLE-BLAP-016√ | 16.548 | $0.41 \pm 0.19$ | – | – | – | [10] |
| OGLE-BLAP-017√ | 25.535 | – | – | – | – | [10] |
| OGLE-BLAP-018√ | 14.022 | $0.40 \pm 0.15$ | – | – | – | [10] |
| OGLE-BLAP-019√ | 48.005 | $2.05 \pm 0.14$ | $28000 \pm 700$ | $4.29 \pm 0.09$ | $-0.76 \pm 0.08$ | [10] |
| OGLE-BLAP-020√ | 47.923 | – | $29200 \pm 500$ | $4.40 \pm 0.07$ | $-0.59 \pm 0.06$ | [10] |
| OGLE-BLAP-021√ | 42.742 | $-5.24 \pm 0.26$ | $28500 \pm 300$ | $4.46 \pm 0.04$ | $-0.64 \pm 0.03$ | [10] |
| OGLE-BLAP-022√ | 74.052 | $6.16 \pm 0.54$ | $28900 \pm 400$ | $4.45 \pm 0.06$ | $-0.74 \pm 0.05$ | [10] |
| OGLE-BLAP-023√ | 19.196 | – | – | – | – | [10] |
| OGLE-BLAP-024√ | 48.268 | $-3.53 \pm 0.26$ | $25200 \pm 300$ | $4.39 \pm 0.05$ | $-0.66 \pm 0.04$ | [10] |
| OGLE-BLAP-025√ | 47.724 | $2.49 \pm 0.74$ | – | – | – | [10] |
| OGLE-BLAP-026√ | 41.250 | $0.27 \pm 0.19$ | – | – | – | [10] |
| OGLE-BLAP-027√ | 48.018 | $0.46 \pm 0.22$ | – | – | – | [10] |
| OGLE-BLAP-028√ | 16.965 | $-5.52 \pm 0.32$ | – | – | – | [10] |
| OGLE-BLAP-029√ | 51.008 | $-4.42 \pm 0.43$ | – | – | – | [10] |
| OGLE-BLAP-030√ | 21.162 | $45.80 \pm 0.65$ | $31400 \pm 300$ | $4.85 \pm 0.05$ | $-0.75 \pm 0.04$ | [10] |
| OGLE-BLAP-031√ | 40.671 | – | $26800 \pm 200$ | $4.38 \pm 0.03$ | $-0.54 \pm 0.03$ | [10] |
| OGLE-BLAP-032√ | 55.968 | $1.98 \pm 0.26$ | – | – | – | [10] |
| OGLE-BLAP-033√ | 15.822 | $0.28 \pm 0.06$ | $33100 \pm 700$ | $5.04 \pm 0.11$ | $-0.88 \pm 0.07$ | [10] |
| OGLE-BLAP-034√ | 44.335 | $2.27 \pm 0.62$ | $30300 \pm 300$ | $4.49 \pm 0.04$ | $-0.62 \pm 0.03$ | [10] |
| OGLE-BLAP-035√ | 28.975 | $1.20 \pm 0.68$ | – | – | – | [10] |
| OGLE-BLAP-036√ | 54.289 | $2.85 \pm 0.66$ | – | – | – | [10] |
| OGLE-BLAP-037√ | 15.712 | – | $32800 \pm 200$ | $4.93 \pm 0.04$ | $-2.15 \pm 0.05$ | [10] |
| SMSS-BLAP-1√(OGLE-BLAP-093√) | 19.5211(20.1679637) | – | $29020^{+193}_{-34}$ | $4.661^{+0.008}_{-0.143}$ | $-2.722^{+0.057}_{-0.074}$ | [11]([12]) |
| OGLE-BLAP-062 √ | 32.9956511 | – | – | – | – | [12] |
| OGLE-BLAP-063 √ | 58.6452979 | – | – | – | – | [12] |
| OGLE-BLAP-064 √ | 40.4852661 | – | – | – | – | [12] |
| OGLE-BLAP-065 √ | 15.1879302 | – | – | – | – | [12] |
| OGLE-BLAP-066 √ | 27.9819187 | – | – | – | – | [12] |
| OGLE-BLAP-067 √ | 20.3038847 | – | – | – | – | [12] |
| OGLE-BLAP-068 √ | 38.3321045 | – | – | – | – | [12] |
| OGLE-BLAP-069 √ | 66.4939433 | – | – | – | – | [12] |
| OGLE-BLAP-070 √ | 34.8664728 | – | – | – | – | [12] |
| OGLE-BLAP-071 √ | 58.4985243 | – | – | – | – | [12] |
| OGLE-BLAP-072 √ | 31.8847179 | – | – | – | – | [12] |
| OGLE-BLAP-073 √ | 36.0712030 | – | – | – | – | [12] |
| OGLE-BLAP-074 √ | 15.3685860 | – | – | – | – | [12] |
| OGLE-BLAP-075 √ | 22.0857709 | – | – | – | – | [12] |
| OGLE-BLAP-076 √ | 34.3470343 | – | – | – | – | [12] |
| OGLE-BLAP-077 √ | 10.1121423 | – | – | – | – | [12] |
| OGLE-BLAP-078 √ | 22.1350836 | – | – | – | – | [12] |
| OGLE-BLAP-079 √ | 41.3926356 | – | – | – | – | [12] |
| OGLE-BLAP-080 √ | 21.6268702 | – | – | – | – | [12] |
| OGLE-BLAP-081 √ | 38.5197428 | – | – | – | – | [12] |
| OGLE-BLAP-082 √ | 53.5543446 | – | – | – | – | [12] |
| OGLE-BLAP-083 √ | 47.7527083 | – | – | – | – | [12] |
| OGLE-BLAP-084 √ | 34.2030899 | – | – | – | – | [12] |
| OGLE-BLAP-085 √ | 46.0804344 | – | – | – | – | [12] |
| OGLE-BLAP-087 √ | 20.4200231 | – | – | – | – | [12] |
| OGLE-BLAP-088 √ | 20.5911629 | – | – | – | – | [12] |
| OGLE-BLAP-089 √ | 25.1626966 | – | – | – | – | [12] |
| OGLE-BLAP-090 √ | 55.4221598 | – | – | – | – | [12] |
| OGLE-BLAP-091 √ | 22.5836372 | – | – | – | – | [12] |
| OGLE-BLAP-092 √ | 7.5140274 | – | – | – | – | [12] |
| OGLE-BLAP-094 √ | 25.3925666 | – | – | – | – | [12] |

**Table 1** – *continued*

| ID | $P$ (min) | $\dot{P}/P$ ($10^{-7}$yr$^{-1}$) | $T_{\rm eff}$ (K) | $\log g$ (c.g.s) | $\log(N_{\rm He}/N_{\rm H})$ | Ref. |
|---|---|---|---|---|---|---|
| | | | special candidates | | | |
| ZTF-sdBV1 | $5.7881 \pm 0.0003$ | – | – | – | – | [8] |
| ZTF-sdBV2 | $6.0096 \pm 0.0003$ | – | – | – | – | [8] |
| ZTF-sdBV3 | $6.0970 \pm 0.0002$ | – | – | – | – | [8] |
| ZTF-sdBV4 | $6.1747 \pm 0.0003$ | – | – | – | – | [8] |
| ZTF-sdBV5 | $6.2615 \pm 0.0003$ | – | – | – | – | [8] |
| ZTF-sdBV6 | $6.2630 \pm 0.0003$ | – | – | – | – | [8] |
| ZTF-sdBV7 | $6.2911 \pm 0.0003$ | – | – | – | – | [8] |
| ZTF-sdBV8 | $7.2577 \pm 0.0003$ | – | – | – | – | [8] |
| ZTF-sdBV9 | $8.7472 \pm 0.0003$ | – | – | – | – | [8] |
| ZTF-sdBV10 | $10.7718 \pm 0.0003$ | – | – | – | – | [8] |
| ZTF-sdBV11 | $11.4748 \pm 0.0003$ | – | – | – | – | [8] |
| ZTF-sdBV12 | $16.5913 \pm 0.0003$ | – | – | – | – | [8] |

*Notes.* Refs.: [1]–P. Pietrukowicz et al. (2017), [2]–T. Wu & Y. Li (2018), [3]–T. Kupfer et al. (2019), [4]–P. R. McWhirter & M. C. Lam (2022), [5]–A. Pigulski et al. (2022), [6]–J. Lin et al. (2022), [7]–G. Ramsay et al. (2022), [8]–T. Kupfer et al. (2021), [9]–J. Borowicz et al. (2023a), [10]–P. Pietrukowicz et al. (2024), [11]–S.-W. Chang et al. (2024), and [12]–J. Borowicz et al. (2023b).

♠ The relative period change rate ($\dot{P}/P$) of targets OGLE-BLAP-002, -004, and -011 are reported by T. Wu & Y. Li (2018, i.e. ref. 2).

♣ The relative period change rate ($\dot{P}/P$) of target ZGP-BLAP-09 is determined by J. Lin et al. (2022, i.e. ref. 6).

‡ HD 133729 and TMTS-BLAP-1 are located in binary systems and are accompanied by a main-sequence B-type star and an unseen object, respectively.

√ and ♡ denote the confirmed targets and the high-confidence candidates, respectively.

⊤ the effective temperature ($T_{\rm eff}$) is an upper limit.

§ $\log(N_{\rm He}/N_{\rm H})$ is estimated from the mass fraction of helium $0.01 < Y < 0.25$ (P. R. McWhirter & M. C. Lam 2022) with the assumption of $Z = 0.02$.

◇ The atmosphere parameters ($T_{\rm eff}$, $\log g$, and $\log(N_{\rm He}/N_{\rm H})$) of OGLE-BLAP-010, -042, -044, and -049 are from P. Pietrukowicz et al. (2024, i.e. ref. 10).

large relative period change rate ($\dot{P}/P = 22.3 \pm 0.9 \times 10^{-7}$ yr$^{-1}$) and short period ($P = 18.9$ min). Distinguishing between the core-helium-burning and shell-helium-burning phases poses a challenge, especially when dealing with the overlapped regions of the evolved core-helium-burning and early shell-helium-burning phases in the $\dot{P}/P - P$ plot, as shown the top panel of Fig. 1. This ambiguity implies that the current analysis cannot definitively excluded the potentially possibility that TMTS-BLAP-1 is an over-evolved core-helium-burning star.

Regarding model constructions, C. M. Byrne & C. S. Jeffery (2018) reported two distinct types of models: pre-White Dwarf (WD) models and pre-Extremely Horizontal Branch (EHB) models with masses of 0.31 and 0.46 $M_{\odot}$, respectively. They investigated the impact of atomic diffusion and radiative levitation on stellar evolution and the pulsating instability of the radial fundamental mode. Their analysis suggested that the lower mass (0.31 $M_{\odot}$) pre-WD models are better than the core H-burning pre-EHB models (0.46 $M_{\odot}$) for interpreting evolutionary speed and pulsating instability. However, as the corresponding models are constriction, with negative relative period change rates, they are inadequate for explaining all the observed period change rates (for further details, see C. M. Byrne & C. S. Jeffery 2018).

C. M. Byrne & C. S. Jeffery (2020) built a set of pre-WD models with masses ranging from 0.18 to 0.46 $M_{\odot}$ by considering the effects of atomic diffusion and radiative levitation. They found that the periods of the OGLE-BLAPs and hg-BLAPs align with the fundamental modes of these models. Their non-adiabatic oscillation analysis indicated that the effective temperature of these models which can excite pulsation reaches up to ∼ 50 000 K. It significantly exceeds the highest effective temperature previously recorded for BLAP targets (∼ 35 000 K). It suggests that more shorter period BLAPs might be discovered in future studies. Furthermore, they also found that high-order p-modes are also excitable for some models.

Recently, H. Xiong et al. (2022) proposed that shell-helium-burning hot sub-dwarf B-type (sdB) stars could explain BLAPs in terms of both period and period changes. On the other hand, A. D. Romero et al. (2018) suggested that BLAPs might be pulsation pre-ELM (Extremely Low-Mass) WD stars with masses of approximately 0.30–0.40 $M_{\odot}$. Their models are stay at core-helium-burning and shell-hydrogen-burning phases. The non-adiabatic pulsation analysis indicates that the pulsation of BLAPs could be non-radial g modes with high radial orders ($l = 1$, $-39 < n < -29$; $l = 2$, $-67 < n < -54$) or low-order radial modes with period range of 20 to 25 min. It is worth noting that the periods of models monotonously decrease along with stellar age (i.e. $\dot{P}/P$ is negative) (for more details, see e.g. fig. 1 of A. D. Romero et al. 2018). Furthermore, A. H. Córsico et al. (2018) partially support the perspective of A. D. Romero et al. (2018), indicating that the $\dot{P}/P$ of non-radial g modes with high radial order are more suitable compared to the radial fundamental modes for explaining the BLAPs. In a separate study, A. H. Córsico et al. (2019) found that among the radial modes, only the fundamental mode is unstable, exhibiting pulsating periods with the range of 20 to 40 minutes, agreeing with observed OGLE-BLAPs periods. However, A. H. Córsico et al. (2019) proposed that interpreting BLAPs pulsating using non-radial high-order g modes might be more suitable than the radial fundamental modes.

Moreover, B. Paxton et al. (2019) took simulations for light-curves and radial-velocity curves utilizing the radial stellar pulsations (RSP; R. Smolec & P. Moskalik 2008) for OGE-BLAP-011, and concluded that the pulsations could be the radial first overtone mode (i.e. 1O-mode) with an approximate mass of 1.0 $M_{\odot}$.

In this study, we aim to conduct a series of non-adiabatic pulsation analyses to elucidate the observational phenomena of BLAPs and explore possible potential phenomena.

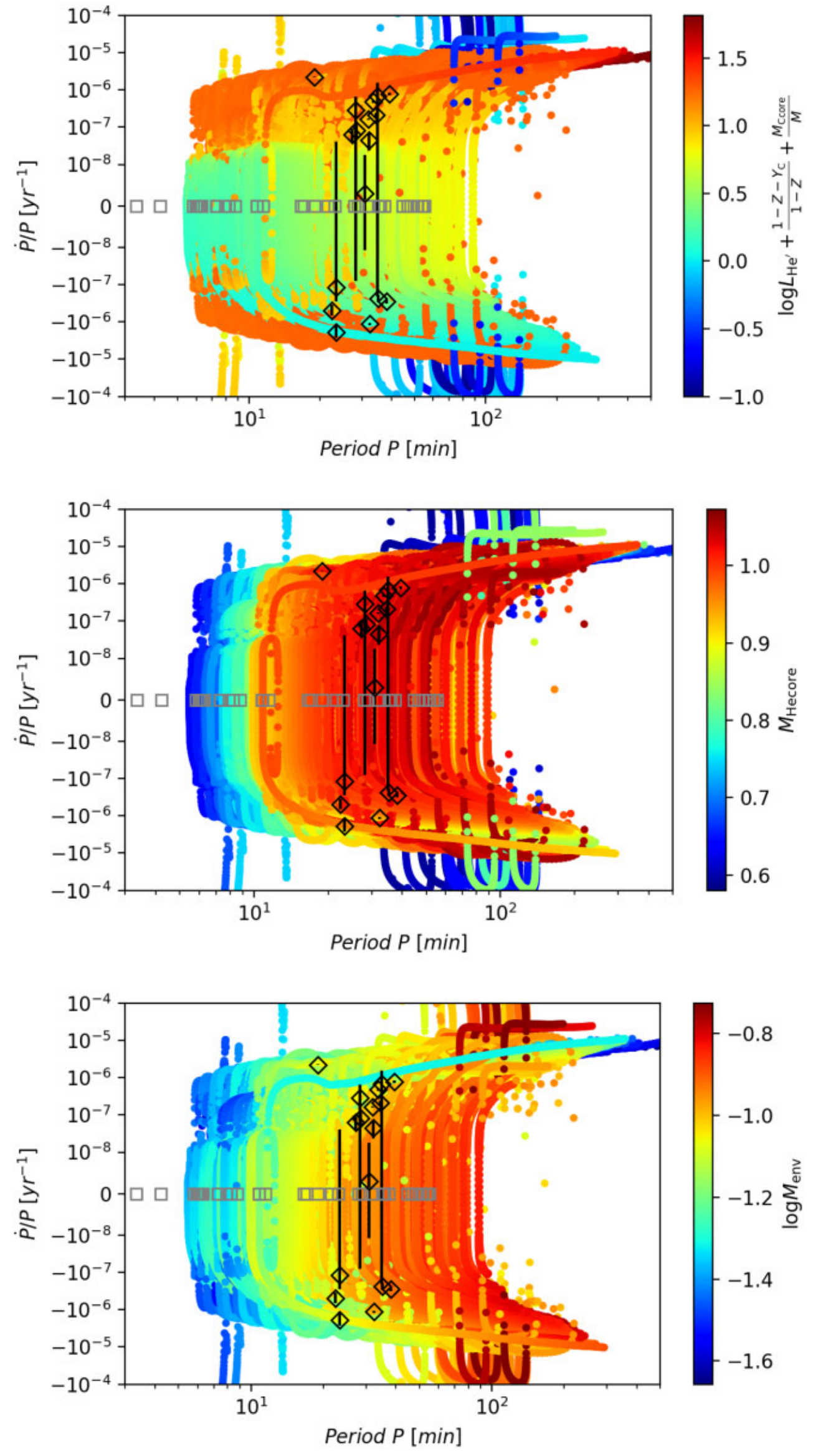

**Figure 1.** Plot of the relative change rate of period $\dot{P}/P$ as a function of pulsating period $P$. The panels, arranged from top to bottom, illustrate different factors: the stellar evolutionary state, the mass of the helium core ($M_{\text{He-core}}$), and the envelope mass ($\log M_{\rm env}$), respectively. The stellar evolutionary state is defined as $S \equiv \log L_{\rm He'} + \frac{1-Z-Y_{\rm C}}{1-Z} + \frac{M_{\text{C-core}}}{M}$, where $Z$ denotes the metal mass fraction, $Y_{\rm C}$ is the centre helium mass fraction, $M_{\text{C-core}}$ represents the mass of the carbon core, and $M$ is the total stellar mass. The term $\log L_{\rm He'} \equiv \log L_{\rm He''}/\min(\log L_{\rm He''})$ is used, with $\log L_{\rm He''} = \log L_{\rm He}(-10 < \log L_{\rm He} < 0)$. Values of $S$ ranging from $-1$ to 0, 0 to 1, and 1 to 2 correspond to stars in the pre main sequence of core-helium-burning, core-helium-burning, and shell-helium-burning phases, respectively. Open diamonds and squares represent the BLAPs with detected or undetected period relative change rates, including candidates (see Table 1). Theoretical models are sourced from T. Wu & Y. Li (2018).

## 2 PHYSICAL INPUTS AND MODEL CALCULATIONS

### 2.1 Physical inputs

In this work, we adopted the Modules for Experiments in Stellar Astrophysics (MESA, v15140) code developed by B. Paxton et al. (2011) to calculate theoretical models. In addition, we utilized the open-source stellar oscillation code, GYRE (the version of gyre-master), based on a new Magnus Multiple Shooting scheme developed by R. H. D. Townsend & S. A. Teitler (2013), to calculate oscillations. Further information on the MESA evolution code can be found in related works by B. Paxton et al. (2013, 2015, 2018, 2019), and about the GYRE oscillation code in the works of R. H. D. Townsend, J. Goldstein & E. G. Zweibel (2018) and J. Goldstein & R. H. D. Townsend (2020).

We adopt the metallicity mixture of GS98 and the OP (M. J. Seaton 2005) opacity table GS98 series (N. Grevesse & A. J. Sauval 1998) denoted as OP-GS98, using the default parameters of the '1M_pre_ms_to_wd' module. We choice the Eddington grey-atmosphere $T-\tau$ relation to describe the stellar atmosphere model, adopt the standard mixing-length theory (MLT) proposed by J. P. Cox & R. T. Giuli (1968) to deal with the convective zone, and fix the mixing-length parameter as $\alpha_{\rm MLT} = 1.9$. Additionally, we adopt the thermonuclear reaction net 'basic_puls_fe56_ni58.net' including the isotopes of $^{56}$Fe and $^{58}$Ni. This is because the pulsation of BLAPs is thought to be excited by the $\kappa$-mechanism due to the accumulation of iron in the metal opacity bump, or Z-bump (for more details, see e.g. P. Pietrukowicz et al. 2017; G. Ramsay 2018; G. Ramsay et al. 2022). More further information on the thermonuclear reaction net can be found in the file path of '$MESA_DIR\data\net_data\nets' (B. Paxton et al. 2011, 2013, 2015, 2018, 2019).

Convective overshooting treatment for the convective core followed the theory of F. Herwig (2000). The overshooting mixing diffusion coefficient $D_{\rm ov}$ exponentially decreases from the outer boundary of the convective core:

$$D_{\rm ov} = D_{\rm conv,0} \exp\left(-\frac{2(z - f_{\rm ov,0}H_{P,0})}{f_{\rm ov}H_{P,0}}\right), \qquad (1)$$

$D_{\rm conv,0}$ and $H_{P,0}$ represent the MLT-derived diffusion coefficient and the corresponding pressure scale height near the Schwarzschild boundary, respectively. Here, $z$ indicates the distance in the radiative layer away from the Schwarzschild boundary; $f_{\rm ov,0}H_{P,0}$ denotes the distance from the convective boundary back into the convective zone; $f_{\rm ov,0}$ and $f_{\rm ov}$ are adjustable parameters (for more detailed descriptions, refer to F. Herwig 2000; B. Paxton et al. 2011, 2013, 2015, 2018). In this work, according to the previous studies by F. Guo & Y. Li (2019, 0.008) for low-mass main-sequence stars and by Z. Li & Y. Li (2021, 0.008) for sdB stars, we set the overshooting parameters of the convective core as $f_{\rm ov} = 0.01$ and $f_{\rm ov,0} = 0.001$.

It is worth noting that the following physics, such as element diffusion, semi-convection, thermohaline mixing, rotation, magnetic field, mass-loss, and radiative levitation, are not included in the theoretical models.

### 2.2 Generating theoretical models

The initial theoretical models used in this research comprise a homogeneous compact He/Z core and a thin H/He/Z envelope. The first step involves creating a full convective He/Z homogeneous model at the Hayashi line, akin to generating a pre-main-sequence model of hydrogen-burning. This is achieved by setting parameters such as 'create_pre_main_sequence_model = .true.', 'initial_mass = 0.35', $Z_{\rm init} = 0.01$, 0.02, and 0.04, and $Y_{\rm init} = 1.0 - Z_{\rm init}$. The model evolves until it reaches the 'zero-age-main-sequence' of helium-burning (denoted by '$\times$' in Fig. 2), resulting in the formation of the initial homogeneous compact He/Z core. The calculations are terminated when the central temperature $\log T_{\rm c}$ reaches 7.8 or the central density log $\rho_{\rm c}$ reaches 4.0. The finalized models at this stage are saved as 'MHecore####Z#####.mod'.

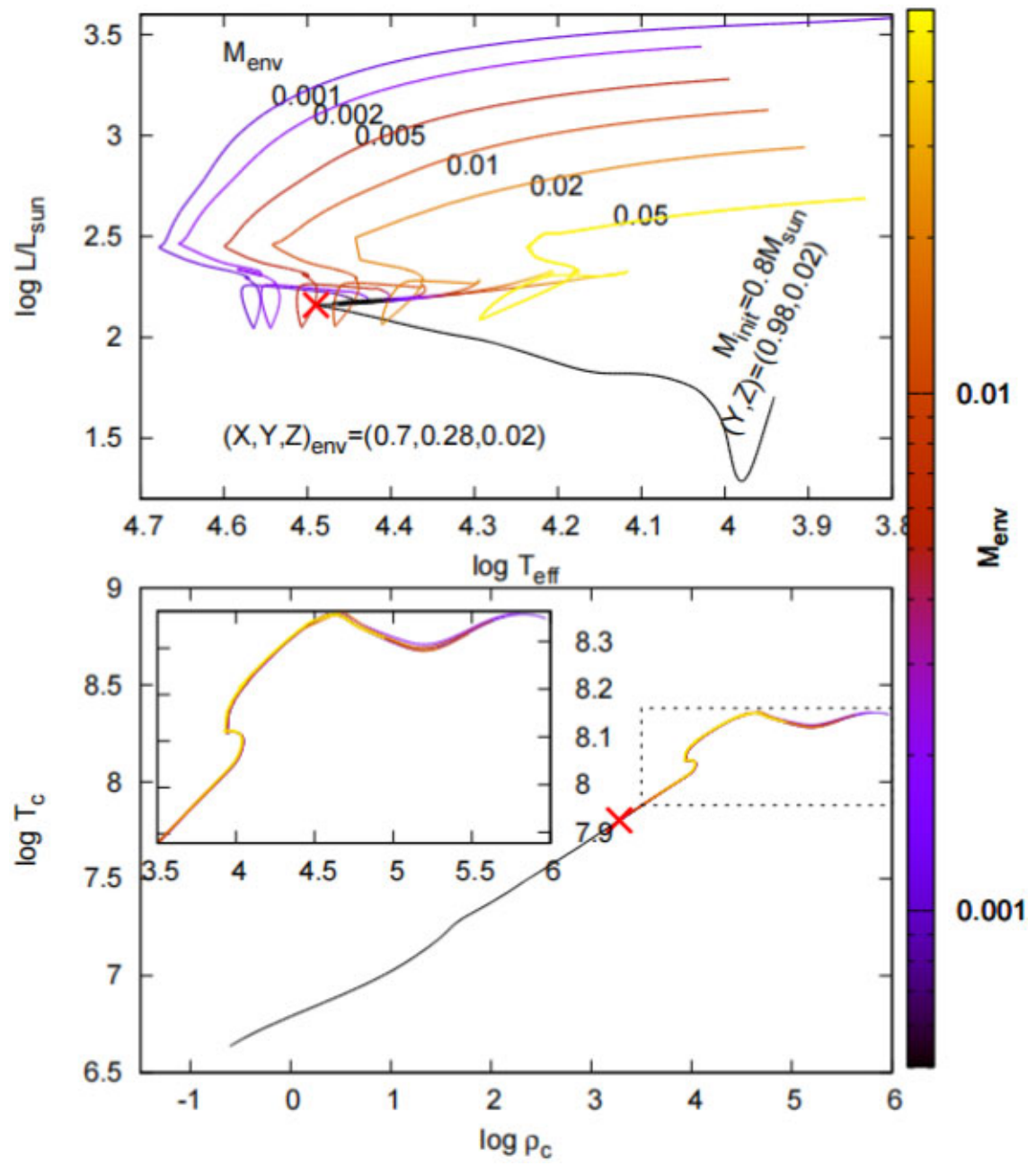

**Figure 2.** $T_{\rm eff}$–$L$ (upper panel) and $\rho_{\rm c}$–$T_{\rm c}$ (lower panel) diagrams show the evolutionary tracks of the models with a helium core of 0.8 $M_{\odot}$ and a series of thin envelopes of 0.001, 0.002, 0.005, 0.01, 0.02, and 0.05 $M_{\odot}$, respectively. The lines are coloured with the palette of envelope mass $M_{\rm env}$. The helium core and thin envelope compositions are $(Y, Z)_{\rm He\text{-}core} = (0.98, 0.02)$ and $(X, Y, Z)_{\rm env} = (0.7, 0.28, 0.02)$, respectively. The symbol '×' indicates the onset of accreting matter.

The subsequent step involves loading the saved model 'MHecore####Z#####.mod' and initiating mass accretion. This process is done by adjusting the mass change parameter (positive values for mass accretion and negative for mass-loss, with 'mass_change = 1E-8') and setting the maximum star mass for mass gain ('max_star_mass_for_gain= $M_{\rm He\text{-}core,\,init} + M_{\rm env,init}$'). The metal mass fraction of the composition of mass accretion (i.e. the thin envelope) is the same as that of the He/Z core, while both He and H elements are adjustable parameters. To ensure completion of mass accretion before helium ignition, we adopt a larger value for the mass change parameter ('mass_change' ranging from $10^{-7}$ to $10^{-6}$ $M_{\odot}$ yr$^{-1}$).

Therefore, we can adjust the following four initial parameters to achieve the model generation process: the mass of the He/Z core ($M_{\rm He\text{-}core,\,init}$), the initial metal mass fraction ($Z_{\rm init}$), the initial hydrogen mass fraction of the envelope ($X_{\rm init}$), and the mass of the envelope ($M_{\rm env,init}$). We set the metal mass fraction at 0.01, 0.02, and 0.04. The other three initial adjustable parameters cover a wide range of values. For instance, $M_{\rm He\text{-}core,init}$ ranges from 0.35 to 1.30 $M_{\odot}$, $M_{\rm env,init}$ ranges from 0.001 to 0.1 $M_{\odot}$, and $X_{\rm init}$ ranges from 0.01 to 0.95. It indicates that the initial helium mass fraction of the envelope varies between 0.98 and 0.04 for $Z_{\rm init} = 0.01$, 0.97 to 0.03 for $Z_{\rm init} = 0.02$, and between 0.95 and 0.01 for $Z_{\rm init} = 0.04$. The detailed distributions of the initial parameters are displayed in Fig. 3.

In addition to the method described above, another approach involves the creation of models through direct evolution. In our previous work, we adopted this method (T. Wu & Y. Li 2018). This process includes the following steps: first, building a pre-main sequence (pre-MS) model at the Hayashi line and evolving to red giant branch (RGB), then removing the outer hydrogen-rich envelope and leaving a compact helium core, and finally adding a thin envelope according to predefined composition. A comparison of the structures of models produced by these two methods reveals that they exhibit similar characteristics, at least within the scope of this study.

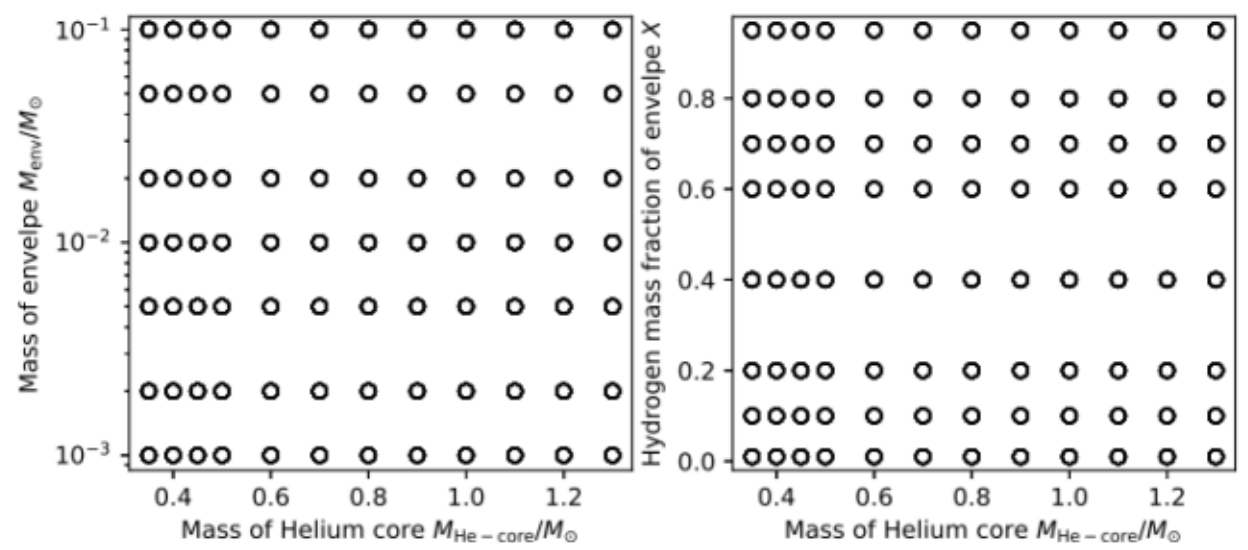

**Figure 3.** Distributions of initial parameters: mass of helium core $M_{\rm He\text{-}core,init} = 0.35 \sim 1.30$ $M_{\odot}$, mass of envelope $M_{\rm env,init} = 0.001 \sim 0.1$ $M_{\odot}$, and the hydrogen mass fraction of outer envelope $X_{\rm init} = 0.01 \sim 0.95$.

## 2.3 Pulsations

We calculated and saved the theoretical structure models and their evolutionary parameters based on the above settings. Subsequently, we utilized GYRE to compute their oscillations, including both the non-adiabatic and adiabatic pulsation, for these theoretical models situated in the range of $3.8 \leq \log T_{\rm eff} \leq 5.0$ and $3.5 \leq \log g \leq 6.5$ on the HR diagram ($\log g$–$\log T_{\rm eff}$) (shown in Fig. 4). For radial modes (i.e. degree $l = 0$), aside from the fundamental mode ($n = 0$), we also computed the other first six low-order modes ($n = 1 \sim 6$). In the case of dipole modes (i.e. degree $l = 1$), we calculated the oscillation modes with periods ranging from 3 to 60 min. Generally, for such models, the frequencies calculated with non-adiabatic and adiabatic theory are nearly identical. None the less, a few models encountered convergence issues when calculate non-adiabatic oscillations. Consequently, we adopted the periods computed with adiabatic theory to determine the period relative change rate $\dot{P}/P$. Additionally, we utilized the imaginary component of the frequencies (Im($\omega$)) calculated from non-adiabatic theory to analyse the pulsation instability.

Generally, the solutions of the wave equation, i.e. the eigenfunctions of oscillations, are expressed in the form of

$$y(x, t) = \tilde{y}(x) \exp(-i\omega t) \quad (2)$$

(details refer to e.g. W. Unno et al. 1989; R. H. D. Townsend 2003; C. Aerts et al. 2010). The frequency $\omega$ is a complex number and expressed as $\omega = \omega_{\rm R} + i\omega_{\rm I}$. The amplitude of oscillations grows (or damps) if the imaginary component of the frequency $\omega_{\rm I}$ is positive (or negative). Correspondingly, the growth rate of oscillation modes is of

$$\eta = \frac{\omega_{\rm I}}{\omega_{\rm R}}. \quad (3)$$

It is also referred to as the unstable coefficient, which characterizes the growth ($\eta > 0$) or damping ($\eta < 0$) rate of the amplitude of oscillation modes over time. Accordingly, the growth time ($\tau_{\rm grow}$) and the damping time ($\tau_{\rm damp}$) of the oscillation mode are defined as:

$$\tau_{\rm grow} = \frac{1}{\omega_{\rm I}} \quad \text{and} \quad \tau_{\rm damp} = -\frac{1}{\omega_{\rm I}}, \quad (4)$$

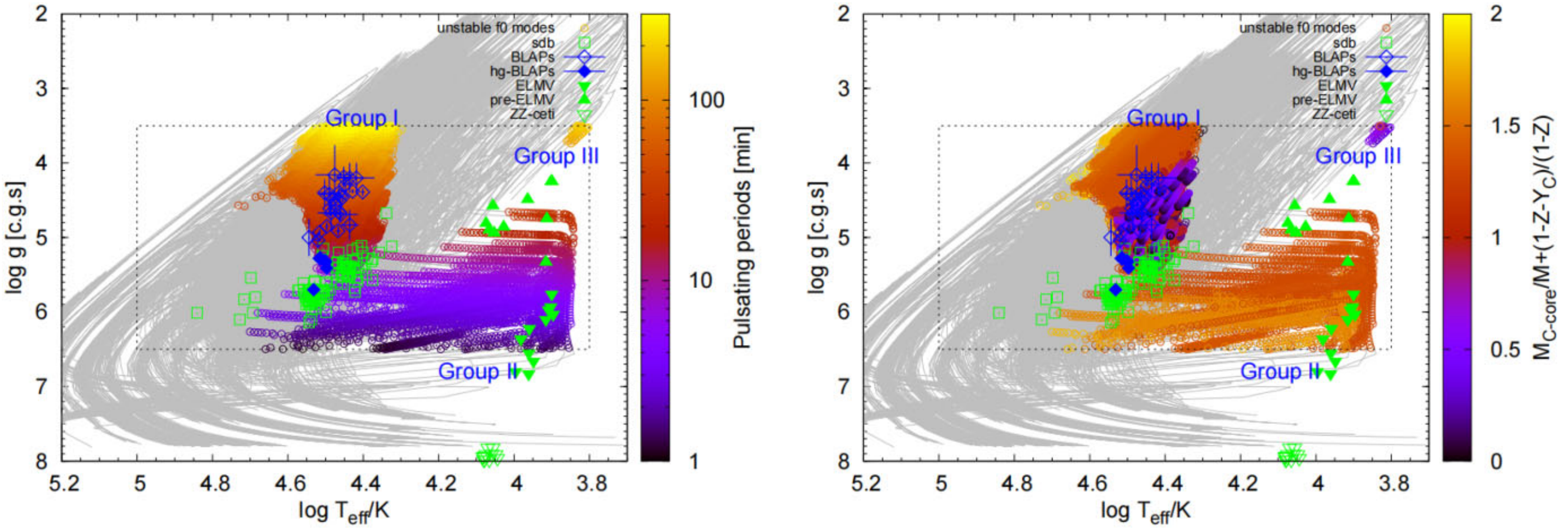

**Figure 4.** $T_{\rm eff}-\log g$, grey line–evolutionary tracks, dashed line–the range of asteroseismically calculated models, open circles–stellar models whose radial fundamental mode ($f_0$) is unstable (i.e. $\eta_{f_0} > 0$), open and filled rhombuses–BLAPs and high-gravity-BLAPs (see Table 1), filled upper and lower triangles–Extremely Low-Mass WDs variable (ELMV) and pre-ELMV (table 8 of A. H. Córsico et al. 2019), open squares–sdBV [collected from D. L. Holdsworth et al. (2017, table 1), J. W. Kern et al. (2018), M. D. Reed et al. (2019), R. Silvotti et al. (2019, 2022), A. S. Baran et al. (2019), S. Charpinet et al. (2019), S. K. Sahoo et al. (2020), G. Ramsay et al. (2022), and R. Jayaraman et al. (2022)], open lower triangles–ZZ Ceti stars (table 1 of G. Fontaine & P. Brassard 2008). Different colours symbolize these models whose $f_0$ are unstable (open circles) according to the palette of the pulsating period ($P$) and stellar evolutionary states ($S$), which are defined as in equation (5), respectively, in the left and right panels.

i.e. the $e$-folding time (see equation 2). In summary, the sign of $\omega_{\rm I}$ (positive or negative) corresponds to the mode instability (unstable or stable). In contrast, the value of $\omega_{\rm I}$ represents how fast the oscillation mode is to grow (or damp) or how easy it is to be excited (or damped).

## 3 RESULTS

Based on the above settings, we calculated more than 2000 theoretical evolutionary tracks. Correspondingly, we calculated their oscillation information (including both adiabatic and non-adiabatic of the radial and dipole modes) for all of these models located in the selected range on the HR diagram, approximately 250 000 models. Here, we will analyse them based on mode excitations, the relative change rate of period $\dot{P}/P$, and the $P-\log g$ relation.

### 3.1 Mode excitations

Pulsations in BLAPs are thought to be excited by the $\kappa$-mechanism due to the Z-bump at the temperature of approximately $\sim 2\times 10^5$ K. Previous studies, such as A. D. Romero et al. (2018), C. M. Byrne & C. S. Jeffery (2018), and C. M. Byrne & C. S. Jeffery (2020), have examined the differential work function (d$W$/d$x$) and the work integral ($W$) of pulsations for both radial and non-radial modes in various stellar models. We will not reiterate these details and will proceed to analyse the calculated results directly.

**The radial fundamental modes:** Fig. 4 shows the HR diagram of the calculated theoretical evolutionary tracks and the models whose radial fundamental modes ($f_0$) can be excited (hereafter, we call them 'unstable models' or 'excitable models'). In the left panel, these models are coloured by the palette of the pulsating period. In the right panel, we introduce a new parameter to present the stellar evolutionary states as a palette:

$$S \equiv \frac{M_{\rm C\text{-}core}}{M} + \frac{1-Z-Y_{\rm C}}{1-Z} \qquad (5)$$

Where $M_{\rm C\text{-}core}$ is the mass of the carbon core, $M$ is the stellar mass, $Y_{\rm C}$ is the central helium mass fraction, and $Z$ is the metal mass fraction. If $S < 1$, helium elements are burning in the core, while if $S > 1$, stars are crossing through the evolved shell-helium-burning phase. Fig. 4 represents that these unstable models are split into three groups. We denoted them as Group I, Group II, and Group III.

Fig. 4 illustrates the distinct characteristics of Group I and Group II stars. Group I exhibits lower gravity ($\log g = 3.5$–$5.6$) and higher effective temperature ($\log T_{\rm eff} = 4.10$–$4.75$), resulting in longer pulsating periods (6–400 min). In contrast, Group II displays higher gravity ($\log g = 4.6$–$6.5$) and lower effective temperature ($\log T_{\rm eff} = 3.8$–$4.7$), leading to shorter pulsating periods (1–30 min). Both groups overlap in the region of pulsating sdB stars.

Group I includes both core- and shell-helium-burning evolutionary phases with masses ranging from 0.4 to 1.4 $M_\odot$. According to theory, the pulsations in these stars are excited by the $\kappa$-mechanism triggered by the ionization of iron and nickel. Notably, apart from the hg-BLAP-1 star with the highest surface gravity among all observed BLAPs, the remaining BLAPs, including normal- and high-gravity BLAPs, can be explained with Group I on the HR diagram. On the other hand, Group II overlaps with the extremely low-mass WD variables (ELMV) and pre-ELMV stars. It corresponds to evolved low-mass shell-helium-burning stars with masses around 0.35 to 0.4 $M_\odot$ ($S > 1$). Correspondingly, the pulsation excitation mechanism for the higher effective temperature models in Group II is similar to that of Group I, involving a $\kappa$-mechanism induced by the Z-bump. For lower effective temperature models in Group II, the pulsations are also excited by the $\kappa$-mechanism, but due to the ionization of H/He elements to enhance opacity. These two groups overlap in the HR diagram ($T_{\rm eff}-\log g$) and are challenging to differentiate. In particular, nearly all these models can excite the radial fundamental mode.

Group III, a few models positioned at the intersection of the lowest effective temperature ($\log T_{\rm eff} \approx 3.8$) and gravity ($\log g \approx 3.5$) exhibit excitable in the radial fundamental mode, as depicted in Fig. 4. These models represent low-mass core-helium-burning stars with a mass range of 0.36 to 0.42 $M_\odot$ and pulsation periods of 130 to 200 min. Their envelope mass varies from 0.01 to 0.02 $M_\odot$, with the hydrogen mass fraction of the envelope ranging between 0.8 and 0.95. Similar to the lower effective temperature models in Group II, the pulsations in these models are also induced the ionization of H/He elements to enhance opacity.

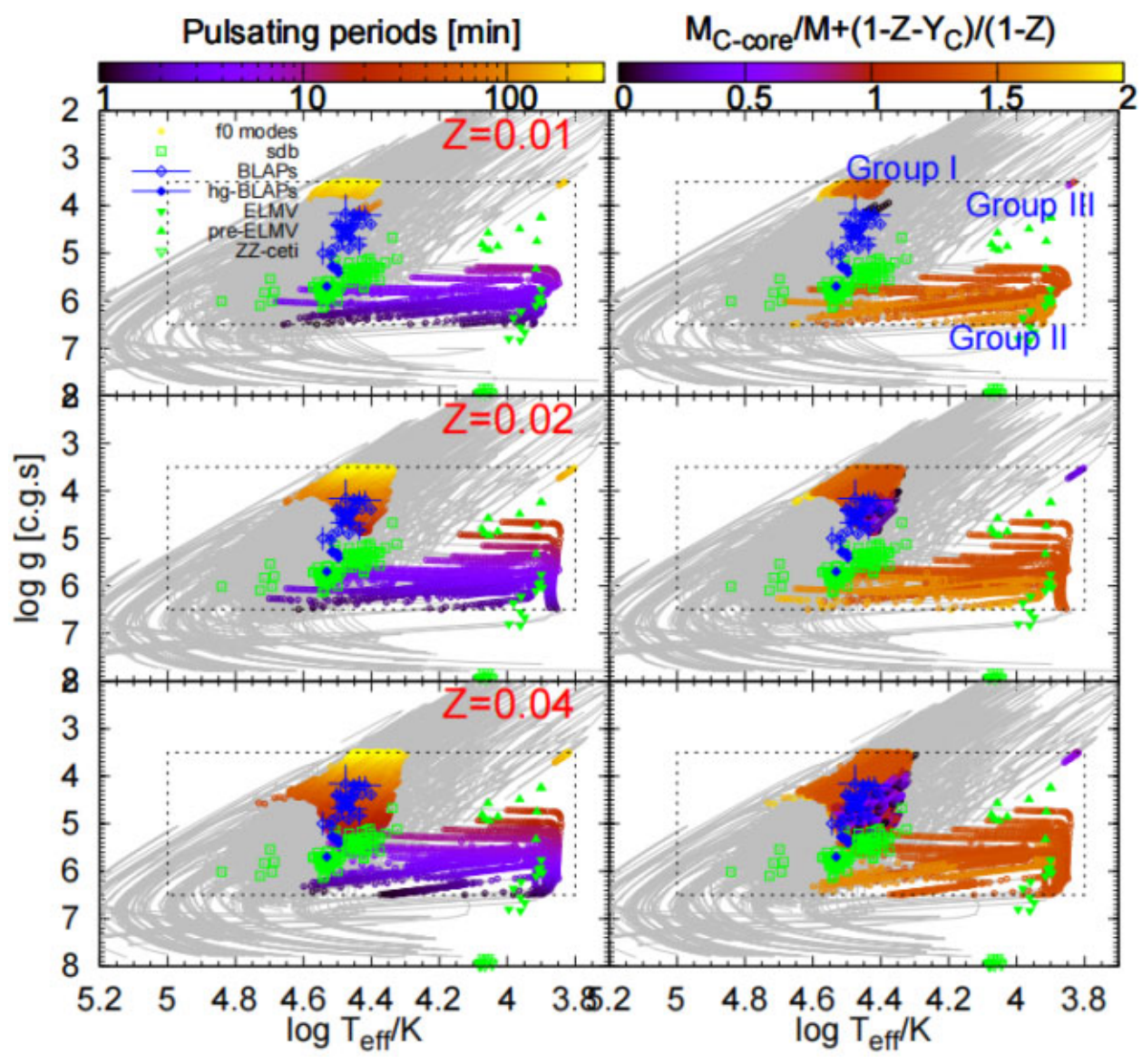

**Figure 5.** Similar to Fig. 4, but divided into three groups with their metal mass fractions $Z = 0.01,\ 0.02$, and 0.04 from top to bottom panels, respectively.

**Impact of metallicity:** Fig. 5 illustrates the influence of metallicity on the excitation of radial fundamental modes. A higher metal content is advantageous for exciting these radial fundamental modes in these models. Notably, among the metal-poor models ($Z = 0.01$), only 16 core-helium-burning models covering the observed BLAPs can excite the radial fundamental modes. These shell-helium-burning phase models are distant from the BLAP areas, suggesting that the metal-poor models only partially explain the observations. The middle panels of Fig. 5 indicate that the solar metal scenario ($Z = 0.02$) provides a more plausible explanation for the BLAPs than the metal-poor scenario. For the metal-rich scenario ($Z = 0.04$), more models can excite the radial fundamental modes. In summary, in the cases of metal-richer models with $Z = 0.02$ and $Z = 0.04$, the number of core-helium-burning models capable of exciting radial fundamental modes increases significantly, with up to 600 models for $Z = 0.02$ and up to 6000 models for $Z = 0.04$. These models exhibit pulsation periods ranging from approximately 10 min to around one hour, covering a wide range of observed BLAPs.

Furthermore, in metal-richer models ($Z = 0.02,\ 0.04$), there is an overlap between core-helium-burning and shell-helium-burning models in the region of BLAPs, making it challenging to differentiate between core-helium-burning and shell-helium-burning phases based on the HR diagram alone. Overall, among the tested metal mass fractions, the metal-richer sets ($Z = 0.04$) are better suited for explaining the observations of BLAPs. This conclusion aligns with the results of adiabatic pulsation analysis presented in T. Wu & Y. Li (2018).

Moving to the ELMV and pre-ELMV region (Group II), the calculated models indicate that nearly all models can excite radial fundamental modes. Their periods range from one minute to tens of minutes. Compared to ELMV and pre-ELMV, these models have higher mass. The varying metal mass fractions mainly impact the distribution of models on the HR diagram in terms of stellar radius, effective temperature, and their corresponding pulsation periods.

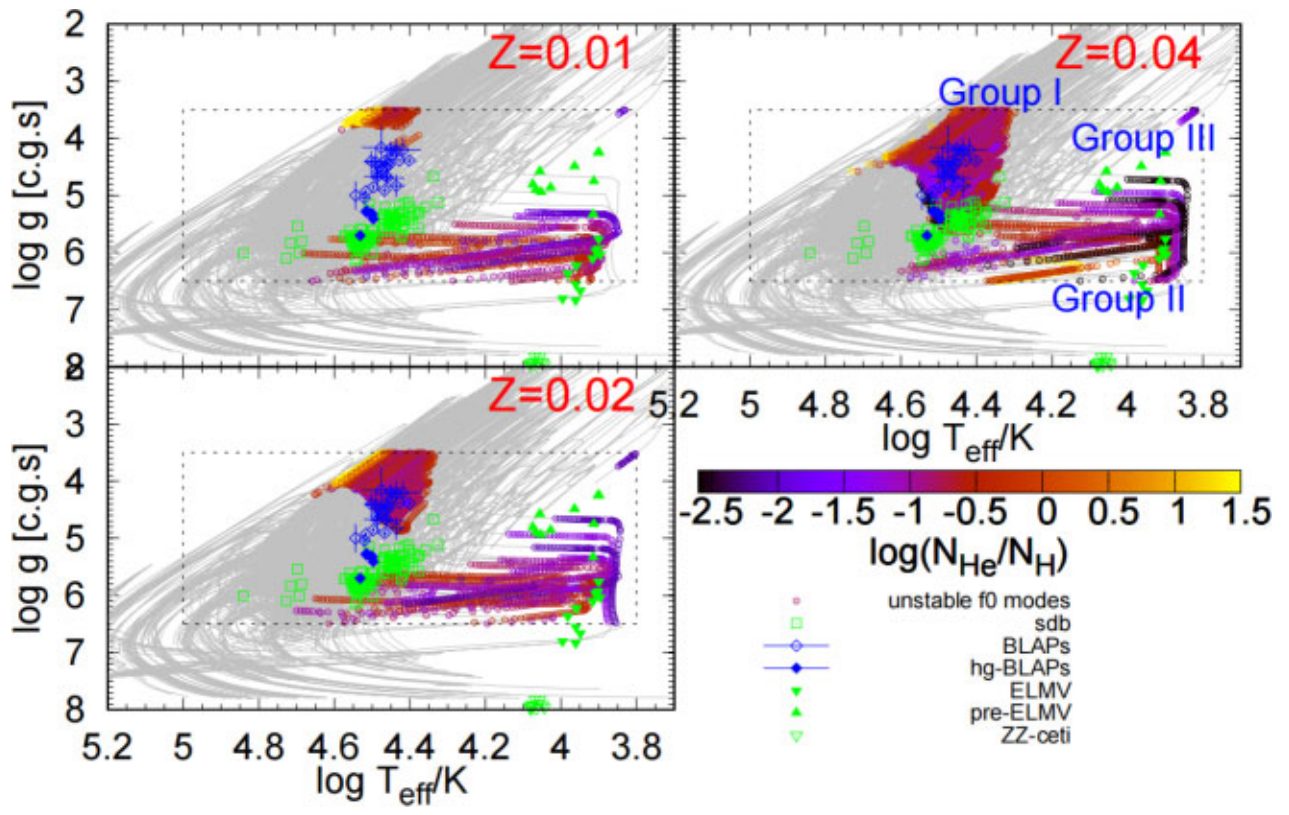

**Figure 6.** Similar to Fig. 5, but the open circles are coloured by the palette of the ratio of element numbers between helium and hydrogen ($\log(N_{\rm He}/N_{\rm H})$).

Conversely, for the lower gravity and cooler models in Group III, changes in metallicity also do not significantly affect the excitation of pulsations or their periods. This reflects that the triggering factors of the excitation mechanism are different from that of Group I.

**The ratio of element numbers between helium and hydrogen ($\log N_{\rm He}/N_{\rm H}$):** The elemental composition of the outer envelopes is characterized by setting the metal mass fraction $Z$ to 0.01, 0.02, and 0.04, with a hydrogen mass fraction $X$ ranging from 0.01 to 0.95. These lead to an initial range of $\log(N_{\rm He}/N_{\rm H})$ values from approximately $-2.0$ to 1.4 for $Z = 0.01$, from about $-2.1$ to 1.4 for $Z = 0.02$, and from around $-2.6$ to 1.4 for $Z = 0.04$. Fig. 6 illustrates that the $\log(N_{\rm He}/N_{\rm H})$ values for models exhibiting unstable radial fundamental modes, similar to observed BLAPs on the HR diagram, vary widely. Specifically, for $Z = 0.01$, the range of $\log(N_{\rm He}/N_{\rm H})$ spans from $-1.5$ to 1.4 in Group I and from $-2.0$ to about 0.0 in Group II. For $Z = 0.02$, the ratio covers the entire parameter space for Group I and ranges from $-2.1$ to approximately $-0.5$ in Group II. Conversely, for $Z = 0.04$, the ratio encompasses the complete parameter space for both Group I and II. Notably, not all evolutionary tracks pass through Group II for different metal mass fractions, implying that the helium-hydrogen ratio in the envelope may not significantly influence the pulsation instability in BLAPs, but influence their radius, effective temperature, and other global parameters.

Table 1 indicates that the observed BLAPs exhibit a ratio ($\log(N_{\rm He}/N_{\rm H})$) ranging from about $-2.8$ to $-0.4$, which corresponds to a mass fraction ratio between helium ($Y$) and hydrogen ($X$) of approximately $Y/X = 0.006$–$1.6$. Theoretical models adequately explain the observations of $\log(N_{\rm He}/N_{\rm H})$. Our model suggests that future discoveries of more helium-rich BLAPs are possible without considering formation channels.

**The thickness of the outer envelope:** For the outer thin envelope mass ($M_{\rm env,init}$), we set it ranging from 0.001 to 0.1 $M_{\odot}$. As shown in Figs 7 and 8, the mass of the outer thin envelope ($M_{\rm env}$) covers a wide range, especially for the Group I, which includes early- and late-evolved stars. Fig. 8 shows that, for most models, due to shell-hydrogen-burning at the outer surface of the helium core, $M_{\rm env}$ progressively decreases along stellar evolution. Particularly, in a few instances of massive helium core models, $M_{\rm env}$ drops below $10^{-4}$ $M_{\odot}$. Noteworthy is that the instability of the radial fundamental modes appears independent of the thickness of the outer thin envelopes of models, being solely influenced by their metal abundances.

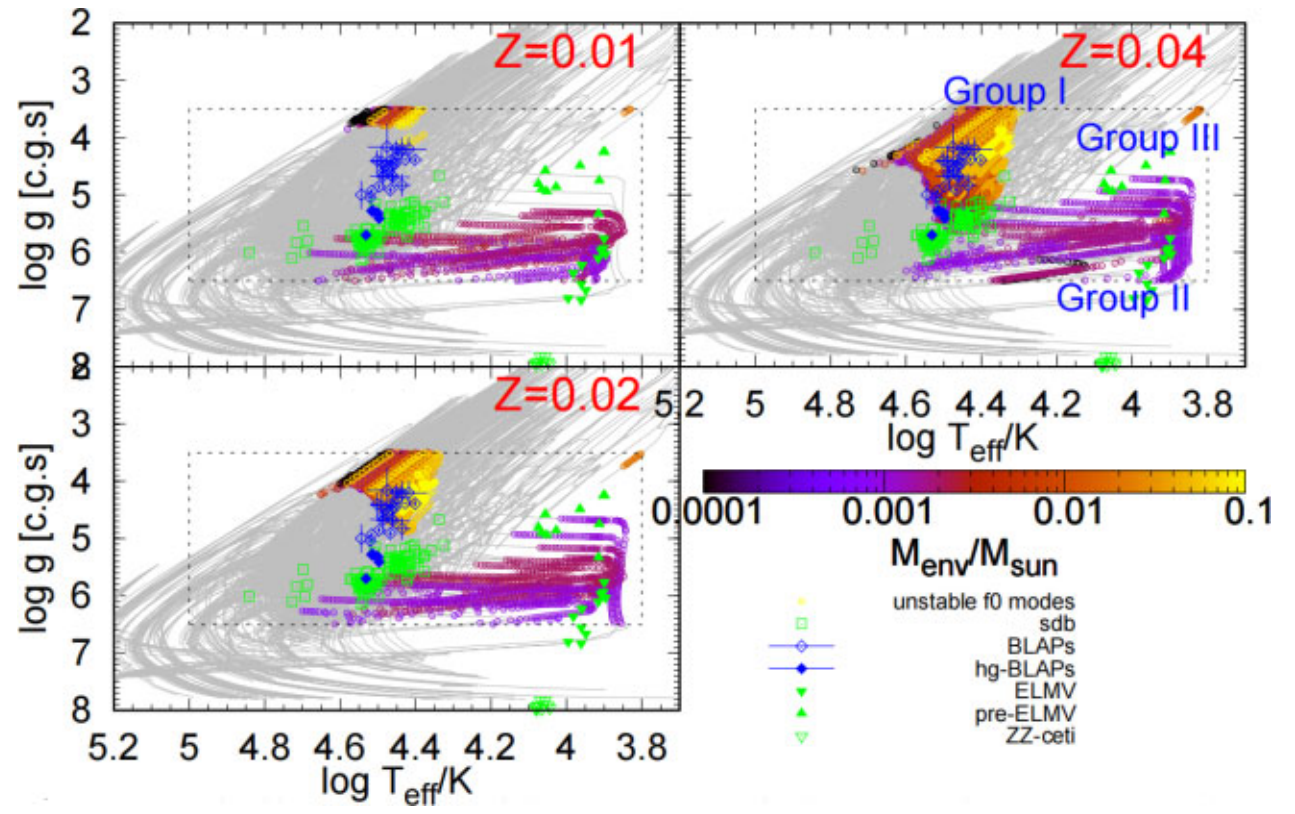

**Figure 7.** Similar to Fig. 6, but the open circles are coulored by the palette of the mass of the outer envelope ($M_{\rm env}$).

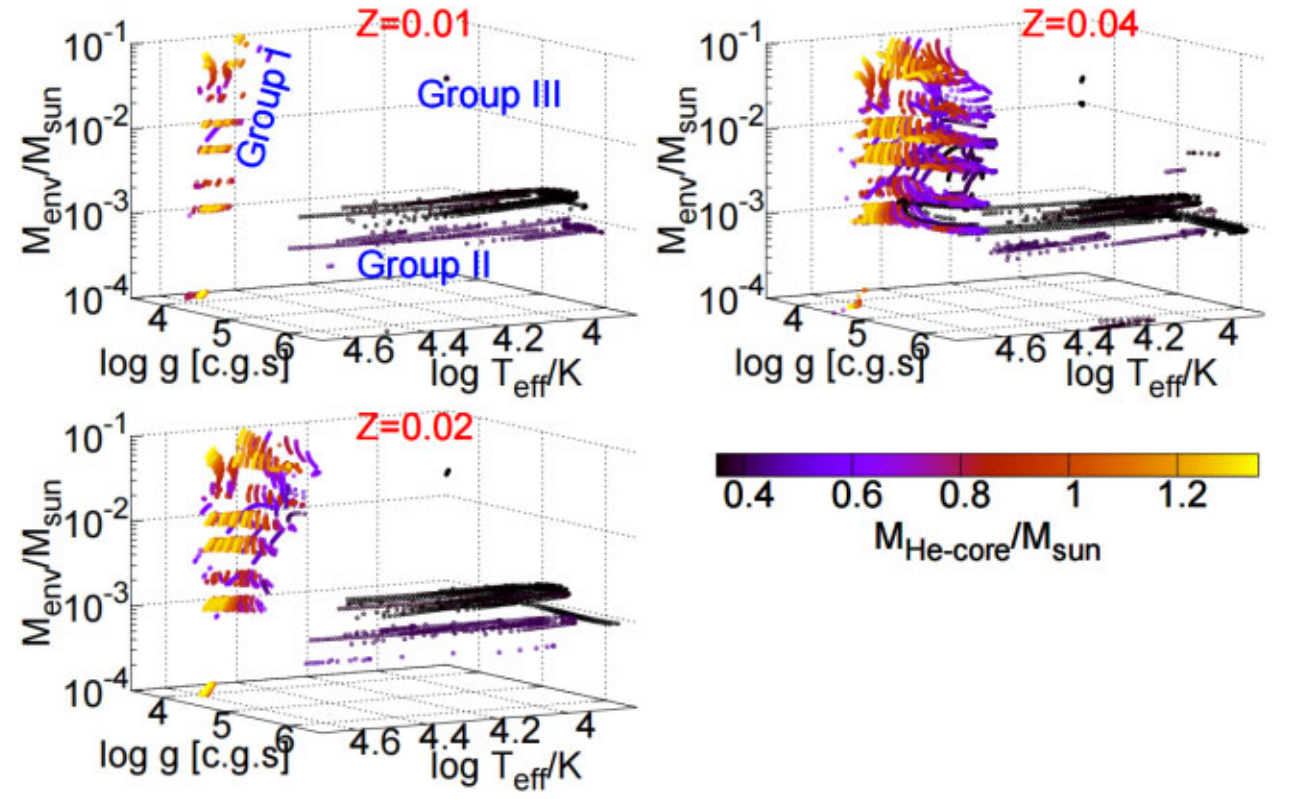

**Figure 8.** The mass of the outer thin envelope $M_{\rm env}$ as a function of $T_{\rm eff}$ and $\log g$. The palette is the mass of the helium core ($M_{\rm He\text{-}core}$). It is noteworthy that $10^{-4}$ is a lower limit of $M_{\rm env}$ for both Fig. 7 and this figure. Lower values are replaced by $10^{-4}$.

Figs 7 and 8 illustrate that, regardless of the metallicity of the models, only those models with a low-mass helium core ($M_{\rm He\text{-}core,\,init} \sim 0.4\ {\rm M}_\odot$) and thin envelope ($M_{\rm env,init} \sim 10^{-3}\ {\rm M}_\odot$) can cross the pre-ELMV or ELMV regions on the HR diagram. We denoted those models as Group II. For the Group III, models have a small helium core ($M_{\rm He\text{-}core,\,init} \sim 0.4\ {\rm M}_\odot$) and a medium envelope ($M_{\rm env,init} \sim 1$–$2 \times 10^{-2}\ {\rm M}_\odot$).

**The first overtone:** The radial first overtone (i.e. $l = 0$ and $n = 1$ mode, denoted as $f_1$) is depicted in Fig. 9. The regions of $f_1$ excitable models are like the radial fundamental modes. They are mainly divided into three groups and also denoted as Groups I, II, and III. Unlike the radial fundamental modes, Groups I and II do not overlap. In the region of sdB, almost no models can excite $f_1$ modes. Except for hg-BLAPs, which is represented by blue full diamonds in the figures, the $f_1$ excitable models also explain all observed normal BLAPs on the HR diagram. Notably, there are three higher gravity hg-BLAPs being over the lower edge of the Group I models.

Compared to the radial fundamental mode, the pulsating instability regions of the radial first overtone are slightly narrower on the HR diagram, particularly at the cooler and lower gravity edges of Group I. However, at the hotter edge of Group I, more evolved models can excite the radial first modes with the characteristic pulsation periods of approximately tens of minutes. For Group II, the excitable region of the radial first overtone is narrower than the radial fundamental modes, yet almost all the cooler edge models are still excitable. In the distribution of Group III, the behaviour is similar to that of the radial fundamental mode.

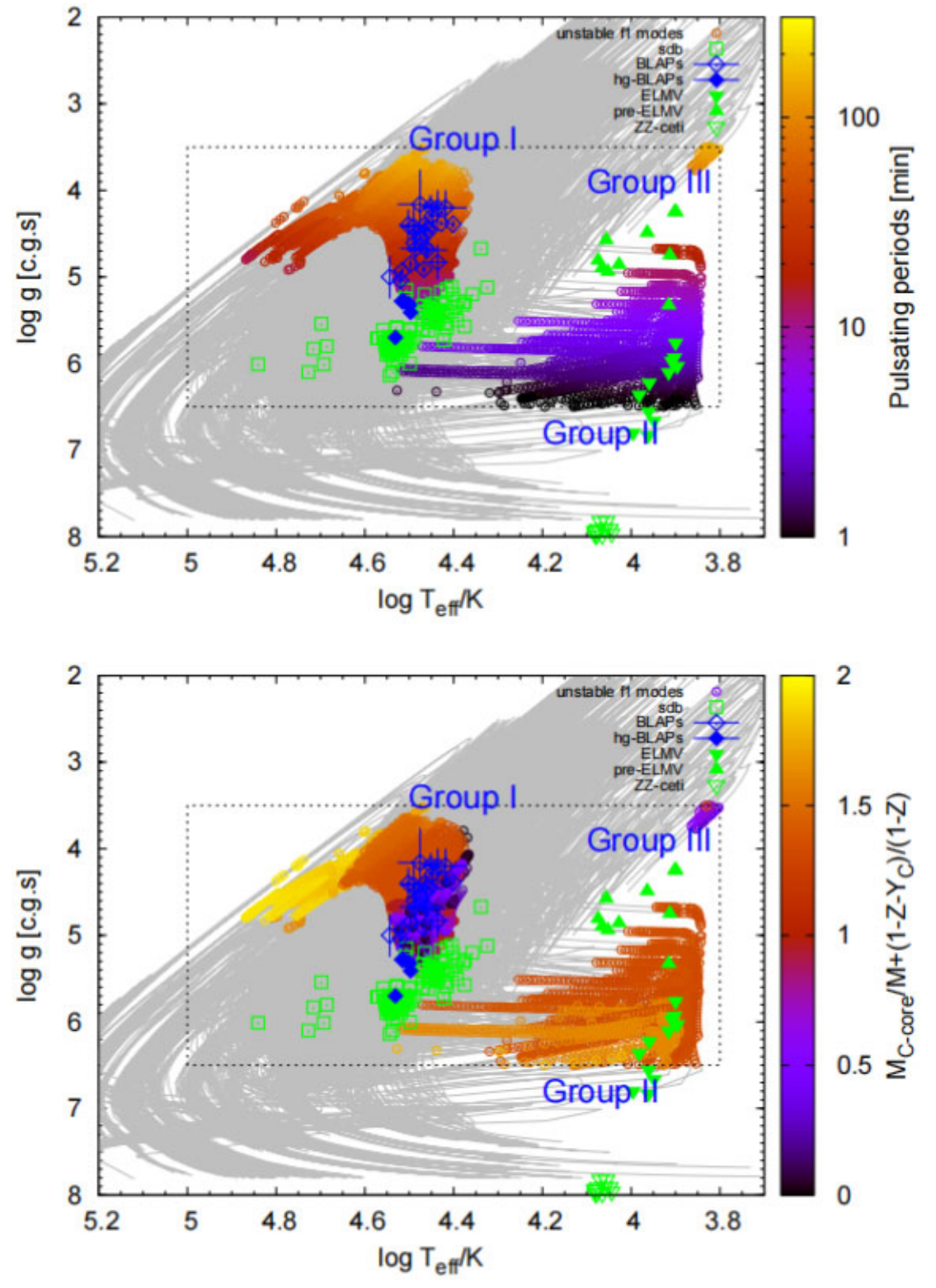

**Figure 9.** Similar to Fig. 4, but for the radial first overtone (i.e. $f_1$).

**Higher order overtone ($f_2$–$f_6$):** Fig. 10 presents the results for the higher-order radial modes ($n = 2$–6). It can be seen from Fig. 10 that, for the second overtone (i.e. radial order $n = 2$ and degree $l = 0$, noted as $f_2$), Group I is divided into two sub-parts. The cooler part is located on the region of normal BLAPs and can be used to partly explain a few normal BLAPs on the HR diagram. In comparison, the hotter part are far away from the normal BLAPs. They are evolved models at the shell-helium-burning evolutionary phase. In Group II, the hotter edge contracts and narrows towards the cooler edge compared to the radial fundamental mode and the radial first overtone. However, the pulsation instability of the radial second overtone is similar to that of the radial fundamental mode and first overtone for the cooler models, indicating that all cooler temperature models in this region can excite the radial second overtone.

It is evident from Fig. 10 that models close to the BLAPs on the HR diagram cannot excite any other higher-order radial modes ($f_3$–$f_6$). Merely a few evolved models can excite these high-order modes. From $f_3$ to $f_6$, the number of models in Group I becomes progressively smaller and smaller, and the corresponding areas are narrower and narrower. Groups II and III show similar trends. Especially, no cooler temperature and lower gravity models can excite $f_6$, leading to the disappearance of Group III. Similarly, in the region of Group II, only a few models can excite $f_6$. In summary, in our models, the excitable areas on the HR diagram (or model

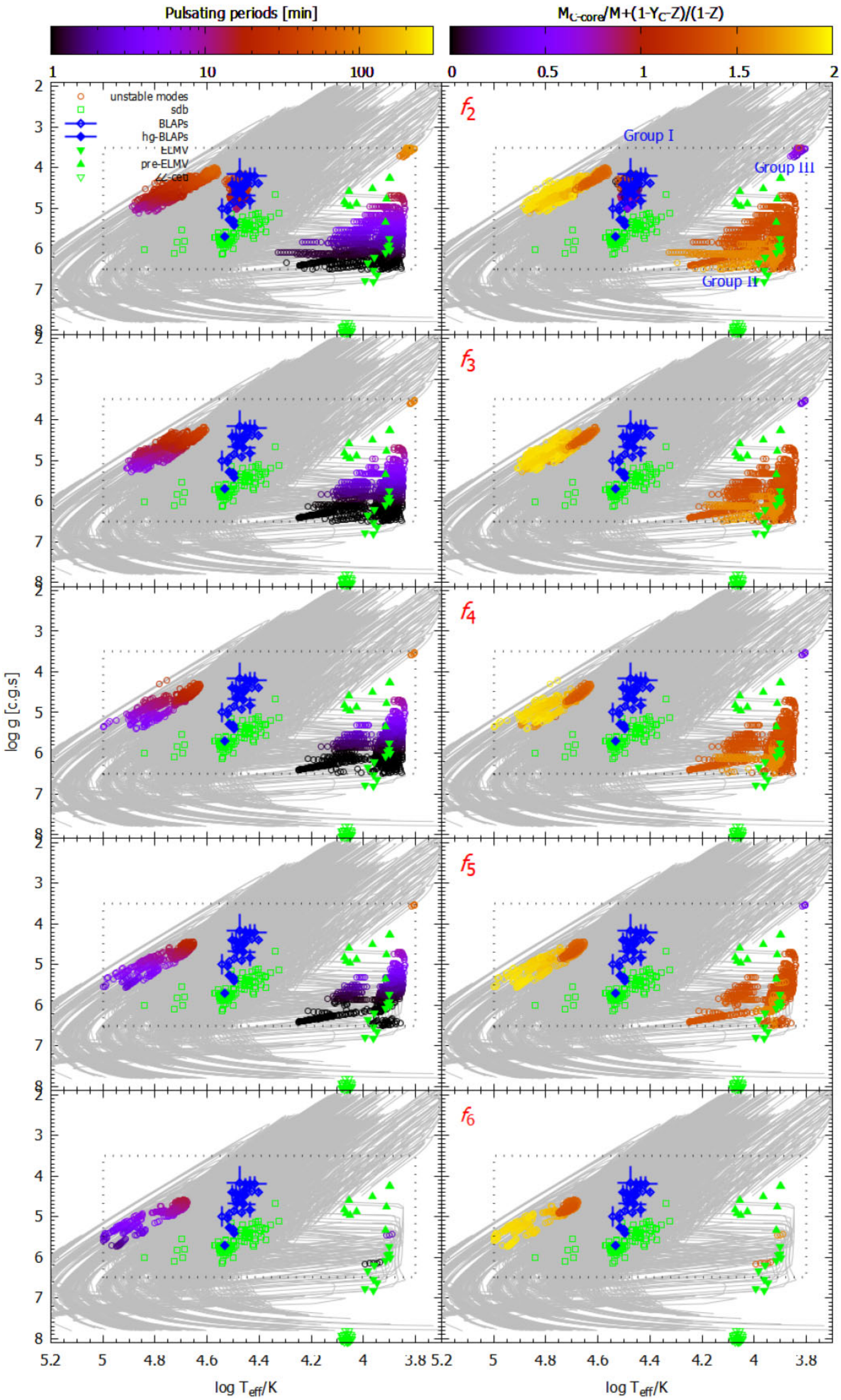

**Figure 10.** Similar to Fig. 4, but for the radial second to sixth overtone ($f_2$–$f_6$) from top to bottom panels.

numbers) decrease with the increase of the radial order for the radial modes.

### 3.1.1 *Dipole modes*

The results of the dipole modes are shown in Figs 11 and 12. In Fig. 11, the unstable models are coloured by the palette of stellar evolutionary states ($S$, see equation 5), the lowest and the highest period of the unstable dipole modes, and the lowest and highest radial order of ones, respectively, from the top to the bottom panels. Fig. 11 represents that these models, whose dipole modes are excitable, are distributed across a wider region compared to those of the radial modes (also refer to Figs 4, 9, and 10). They can be found anywhere, even though some areas are scarce. They also mainly concentrate on three large regions, but correspond to the BLAPs, sdB, and (pre-) ELMV on the HR diagram. For the group of BLAPs, its distribution shape is similar to the Group I of the radial fundamental mode and the first overtone (see Figs 4 and 9).

Compared with the radial fundamental modes (refer to Fig. 4), these lower surface gravity ($\log g = 3.5$–$4.2$) models almost cannot excite dipole modes with the period of $P = 3$–$60$ min. Only a few models can excite a few dipole modes. Therefore, this is not conducive to explaining the current observed BLAPs.

It is noteworthy that, at the upper edge of the sdB region, most of the models that can excite dipole modes with periods between 3 and 60 min are in the core-helium-burning phase (see the top panel of Fig. 11). It seems beneficial to explain the excitation of the pulsation of sdB stars.

From the middle two panels of Fig. 11, it can be observed that in the corner of high surface gravity and cooler temperature (i.e. pre-ELMV and ELMV areas), only a few shorter-period ($P < 15$ min) dipole modes can be excited; almost all the long-period dipole modes cannot be excited in this area. The bottom two panels further show that the corresponding radial order $n_{pg}$ ranges $-20 \sim 2$.

Fig. 12 displays the distributions of the periods and radial orders for the highest and the lowest period dipole modes. In the top panel of Fig. 12, a multitude of models have a narrow excitable frequency range, especially for these models that are in the core-helium-burning phase, with the lowest periods of the excitable modes ranging from 10 to about 25 min. Only a few evolved models have a wider excitable frequency range. For all unstable dipole modes, the radial orders $n_{pg}$ mainly concentrate on $-10 \sim 2$ as shown in the bottom panel of Fig. 12.

It is worth noting that in the region of BLAPs, both the lowest and highest periods of excitable modes range across 3 to 60 min and almost monotonously increase with the decrease of the surface gravity (see Fig. 17). These imply that, in this region, the excitable dipole modes distribute into a narrow band. The corresponding radial order $n_{pg}$ ranges from about 1 to $-200$ with the decrease of surface gravity, and mainly concentrates on 0.

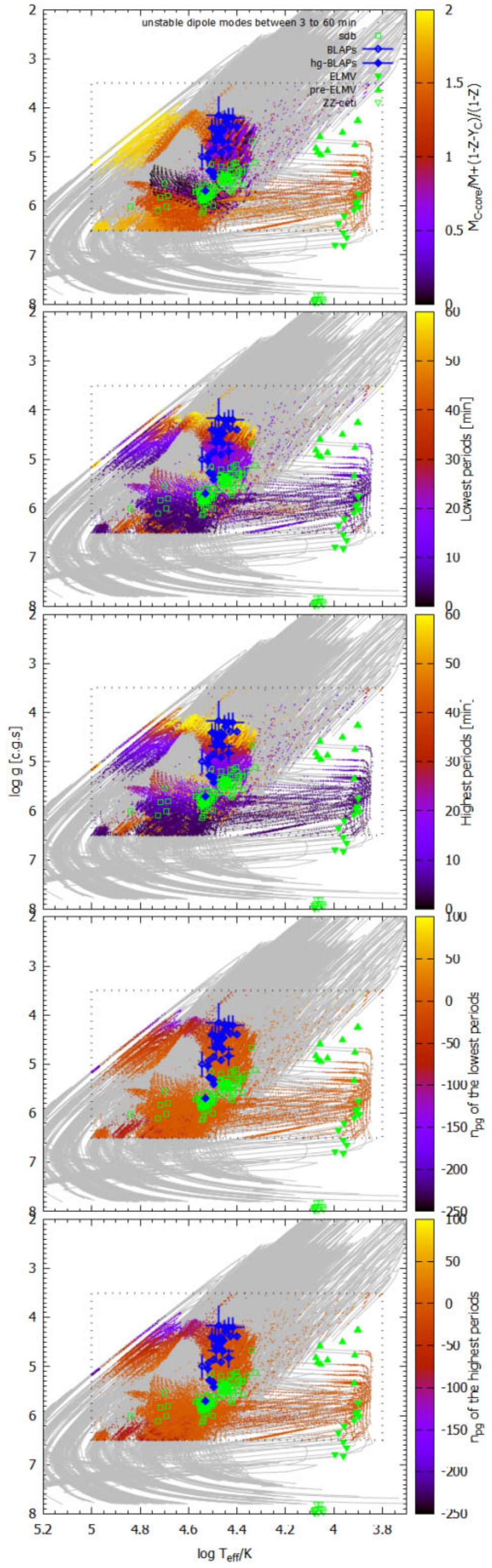

**Figure 11.** Similar to Fig. 4, but for the dipole modes ($l = 1$) with periods ranging from 3 to 60 min. From the top to the bottom panels, the colour palettes are stellar evolutionary states ($S$, see Fig. 4), the shortest and longest periods, and the lowest and highest radial order of the unstable dipole modes, respectively.

### 3.1.2 *The other opacity tables*

In the above sections, we present the results using the metal composition of N. Grevesse & A. J. Sauval (1998, abbreviated as GS98) and the opacity table of M. J. Seaton (2005, abbreviated as OP), denoted as OP-GS98. We conduct comparisons among various combinations of different metal compositions, which include GN93

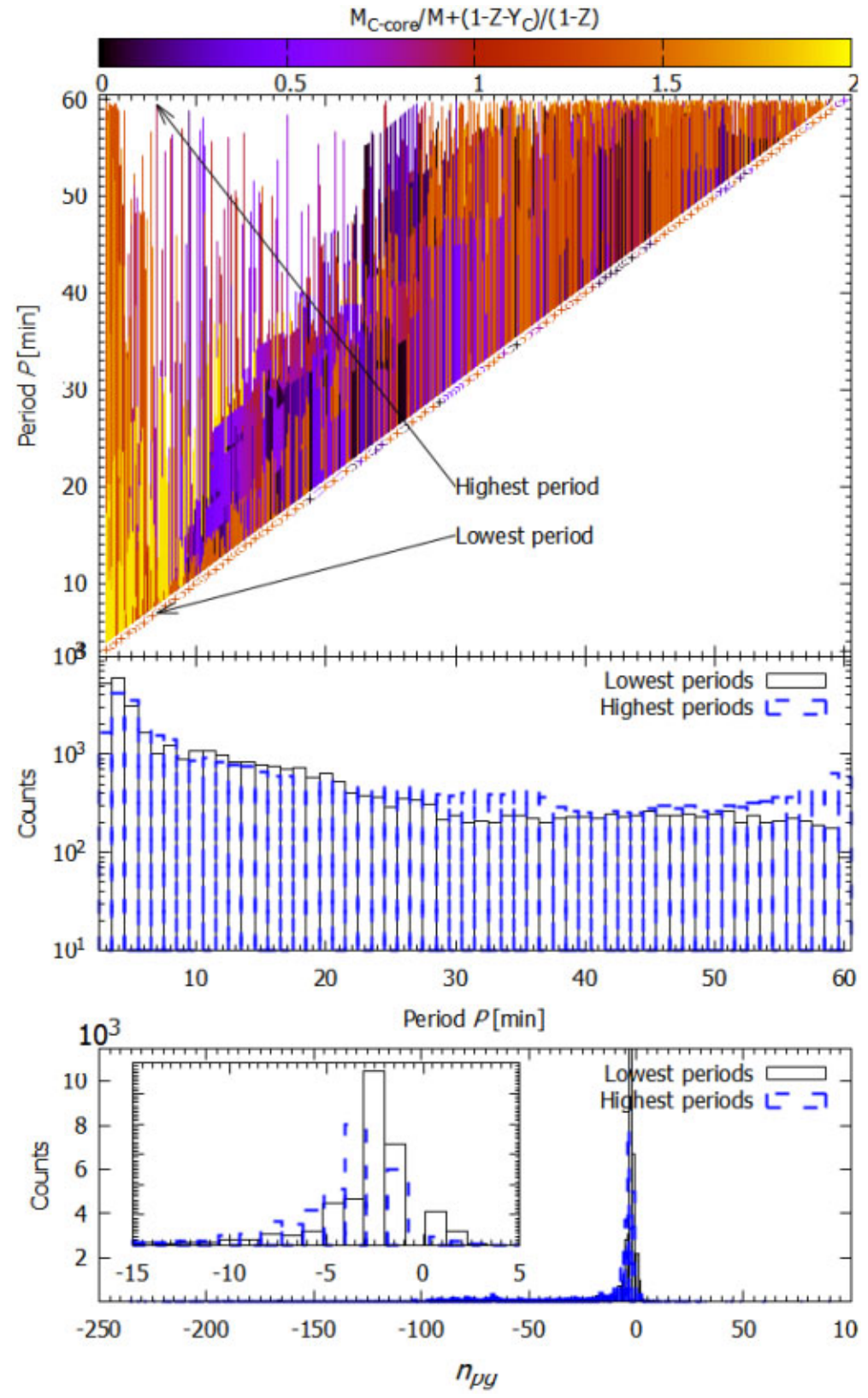

**Figure 12.** The period ranges of the unstable dipole modes are shown as a function of the lowest period in the upper panel, and the statistical distribution of the lowest periods (black solid line) and the highest periods (blue dashed line) in the middle panel. The bottom panel displays the statistical distribution of the radial order of the unstable dipole modes. The solid black line represents the lowest periods, while the blue dashed line denotes the highest periods. In the upper panel, these lines are colour-coded based on the palette of stellar evolutionary states $S$ (see Fig. 4).

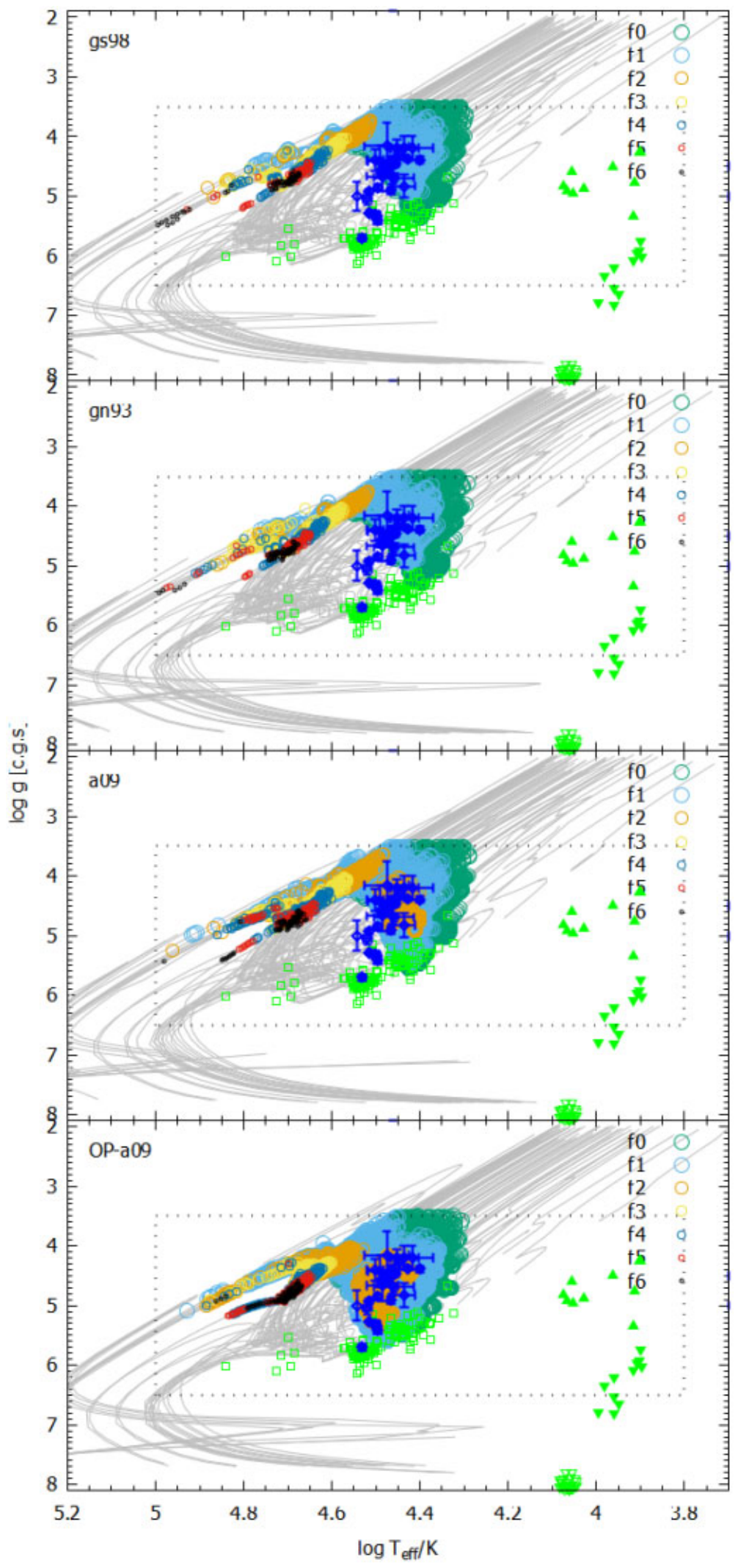

**Figure 13.** Similar to Fig. 4, but for the other different combinations between different opacity table series (OP and OPAL) and different metal compositions (GS98, GN93, and A09). The previous analysis combines the OP table and the GS98 composition, i.e. OP-GS98. Here, the combinations of GN93 (OPAL + GN93), GS98 (OPAL + GS98), A09 (OPAL + A09), and OP-A09 (OP + A09) are shown from top to bottom, respectively. The corresponding initial parameter spaces are $Z_{\rm init} = 0.04$, $M_{\rm He\text{-}core,init} = (0.5, 0.7, 0.9, 1.1)\,{\rm M}_\odot$, $M_{\rm env,init} = (0.001, 0.005, 0.01, 0.05, 0.1)\,{\rm M}_\odot$, and $X_{\rm init} = (0.01, 0.1, 0.4, 0.6, 0.8)$. The unstable modes of $f_0$–$f_6$ are displayed in various colours.

(N. Grevesse & A. Noels 1993), GS98 (N. Grevesse & A. J. Sauval 1998), and A09 (M. Asplund et al. 2009), and different opacity table series, namely OP (M. J. Seaton 2005) and OPAL (C. A. Iglesias & F. J. Rogers 1996). Specifically, the other combinations considered are GN93 (OPAL+GN93), GS98 (OPAL + GS98), A09 (OPAL + A09), and OP-A09 (OP + A09). The corresponding initial parameters consist of $Z_{\rm init} = 0.04$, $M_{\rm He\text{-}core,init} = (0.5, 0.7, 0.9, 1.1)\,{\rm M}_\odot$, $M_{\rm env,init} = (0.001, 0.005, 0.01, 0.05, 0.1)\,{\rm M}_\odot$, and $X_{\rm init} = (0.01, 0.1, 0.4, 0.6, 0.8)$. The remaining calculation settings are consistent with those outlined in Section 2.2. The results are shown in Fig. 13.

In comparison with the OPAL opacity table, OP is more effective in interpreting the observations of BLAPs, as demonstrated by A09 and OP-A09 (illustrated in the bottom two panels of Fig. 13). Concerning metal compositions, A09 exhibits superior performance compared to GS98 and GN93 in representing the observations on the HR diagram. Moreover, GS98 and GN93 exhibit nearly equivalent effects, as indicated in the top three panels of Fig. 13. Therefore, among the five combinations considered, OP-A09 emerges as the optimal choice for elucidating the observations of BLAPs. Regarding OP-A09, aside from the highest gravity hg-BLAP (i.e. hg-BLAP-1), almost all the observations can be perfectly covered by both of the unstable radial fundamental modes ($f_0$) and first overtone ($f_1$). Correspondingly, models in the BLAPs region on the HR diagram do not promote the excitation of higher-order modes. These high-order modes ($f_3$–$f_6$) are only partially excited in hotter and more evolved models.

### 3.2 The relative period changes

Period change is one of the most important oscillation parameters, which mainly implies the variation of stellar size, i.e. stellar radius, if we ignore the non-linear effect. A positive period change indicates that stars expand, while a negative change corresponds to contraction. Therefore, we can use the period changes to constrain the stellar evolutionary speed. To date, 35 BLAPs have had their period changes measured, as shown in Table 1. In earlier studies, T. Wu & Y. Li (2018) and J. Lin et al. (2022) utilized period change as a significant constraint to determine the evolutionary states of BLAPs.

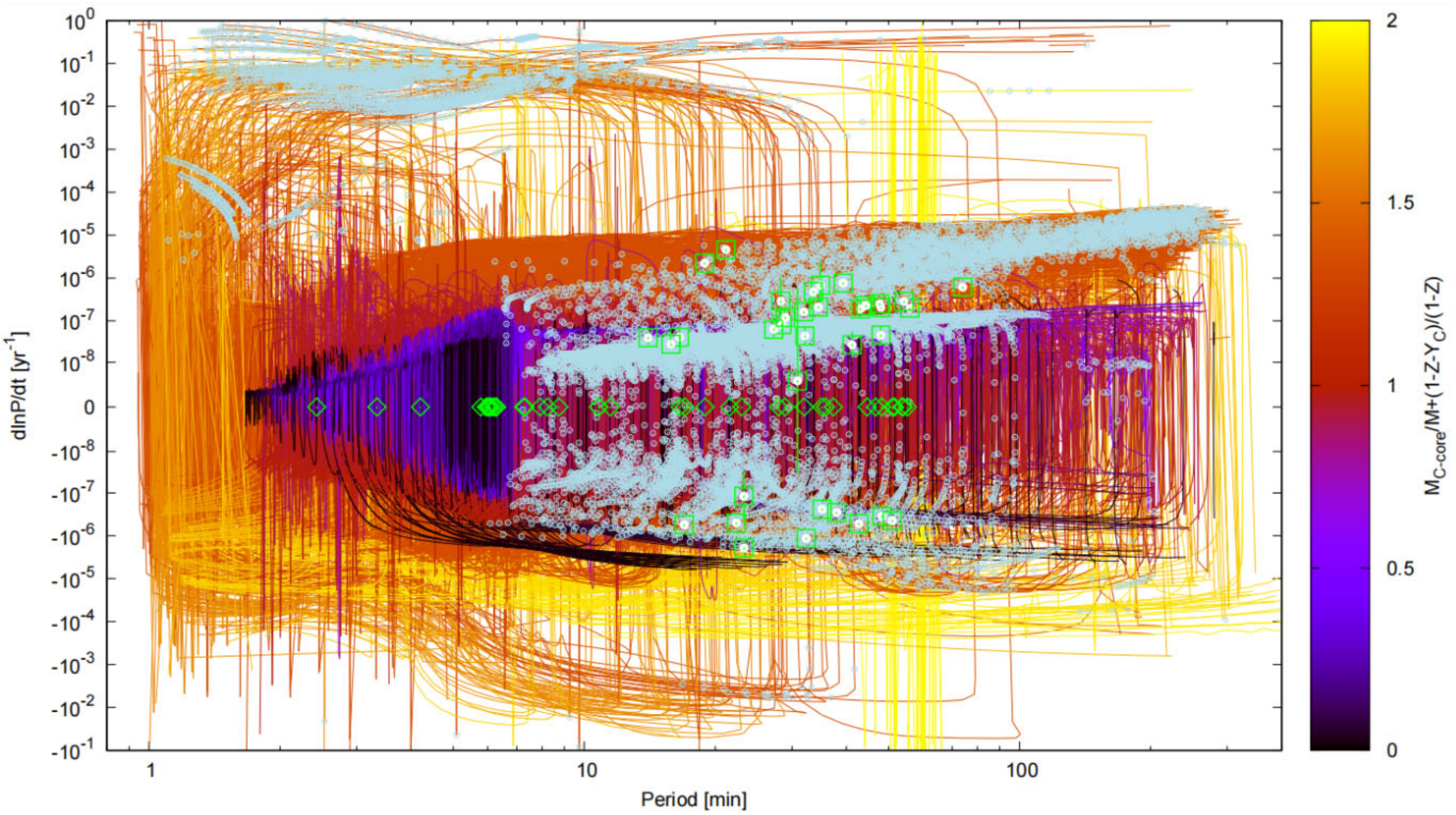

**Figure 14.** The period relative change rate $\dot{P}/P$ as a function of period $P$ for the radial fundamental modes. Open squares: observed BLAPs whose $\dot{P}/P$ are detected, open rhombus: the other observed BLAPs, including normal BLAPs, hg-BLAPs, and candidates. Lines are coloured by the palette of stellar evolutionary states ($S$, see Fig. 4). The models of pulsating instability of the radial fundamental mode are marked with light-blue small open circles.

This method is also employed in high-amplitude $\delta$ Scuti stars to assess their evolutionary states or speeds (see for example J.-S. Niu & H.-F. Xue 2022; H.-F. Xue, J.-S. Niu & J.-N. Fu 2022). Among these 35 BLAPs, 10 exhibit negative period changes, while 25 show positive values. The corresponding period ranges from approximately 14 to 74 min. OGLE-BLAP-030 exhibits the largest period change ($\dot{P}/P = 45.80 \pm 0.65$ yr$^{-1}$), indicating that it is the fastest evolving star. TMTS-BLAP-1 (also known as ZGP-BLAP-01) follows as the second fastest ($\dot{P}/P = 22.3 \pm 0.9$ yr$^{-1}$) and is considered as a shell-helium-burning star (for more details, see J. Lin et al. 2022).

The period changes of the radial fundamental modes are depicted in Fig. 14 (called as $P - \dot{P}/P$ plot). Similar to the previous figures, the model calculations are coloured with the value of $S$, which presents stellar evolutionary states. The light-blue open circles denote these models which can excite the radial fundamental mode. Observations are represented by green rhombuses, if the period change is undetected, and squares.

From Fig. 14, it is evident that, for most models, the period changes fall within $10^{-5}$ yr$^{-1}$. A few models exhibit larger changes (up to $10^{-1}$ yr$^{-1}$) and lower period (down to $\sim$ 1 min). These models mainly correspond to ones of Group II of Fig. 4. Theoretical models align well with current observations for BLAPs with detected period changes. The figure illustrates that core-helium-burning stars have lower period changes compared to those shell-helium-burning stars. The former within $-10^{-5} \sim 10^{-6}$ yr$^{-1}$ and the latter extend to $-10^{-3} \sim 10^{-4}$yr$^{-1}$. Notably, most of the late-evolved excitable models have positive period changes. Fig. 14 also shows that most BLAPs are core-helium-burning stars. Only a few potentially are shell-helium-burning stars. Surely, clearly distinguishing whether they are late-evolved core-helium-burning or shell-helium-burning stars is challenging as they overlap in the $P - \dot{P}/P$ plot and the HR diagram.

Fig. 14 indicates that models with positive period changes outnumber those with negative changes. In observations, the proportion of BLAPs with positive period changes exceeds two-thirds, while those with negative values are less than one-third. The distribution of the period changes suggests that the observations accurately reflect the true nature of BLAPs without any observational selection bias. Importantly, the unstable modes mainly reside in the period range of 6300 min. For lower period models (1–6 min), the unstable modes with larger period changes ($\dot{P}/P > 10^{-6}$ yr$^{-1}$) are excited in the late-evolved shell-helium-burning models. For these models, except their period, the other observable parameters, for instance $T_{\rm eff}$ and $\log g$, are disagree with hg-BLAPs. In contrast, the smaller period change modes cannot be triggered in the core-helium-burning and early-evolved shell-helium-burning models. Further observations are needed to validate these findings.

### 3.3 Period–log $g$, period–log $R$, and period–$\bar{\rho}$ relations

For a star, the pulsating period ($P$) and the mean density ($\bar{\rho}$) are related by the basic pulsation relation

$$Q = P\sqrt{\bar{\rho}/\bar{\rho}_{\odot}} \quad (6)$$

(refers to e.g. M. Breger 1979; M. Breger et al. 1999; M. Breger & M. Montgomery 2000), where $Q$ is called the pulsation constant, which varies with stellar evolution. For a specific mode, $Q$ is almost a constant for all $\delta$ Scuti stars. Therefore, it is usually used to determine the mean stellar density of $\delta$ Scuti stars (refers to e.g.

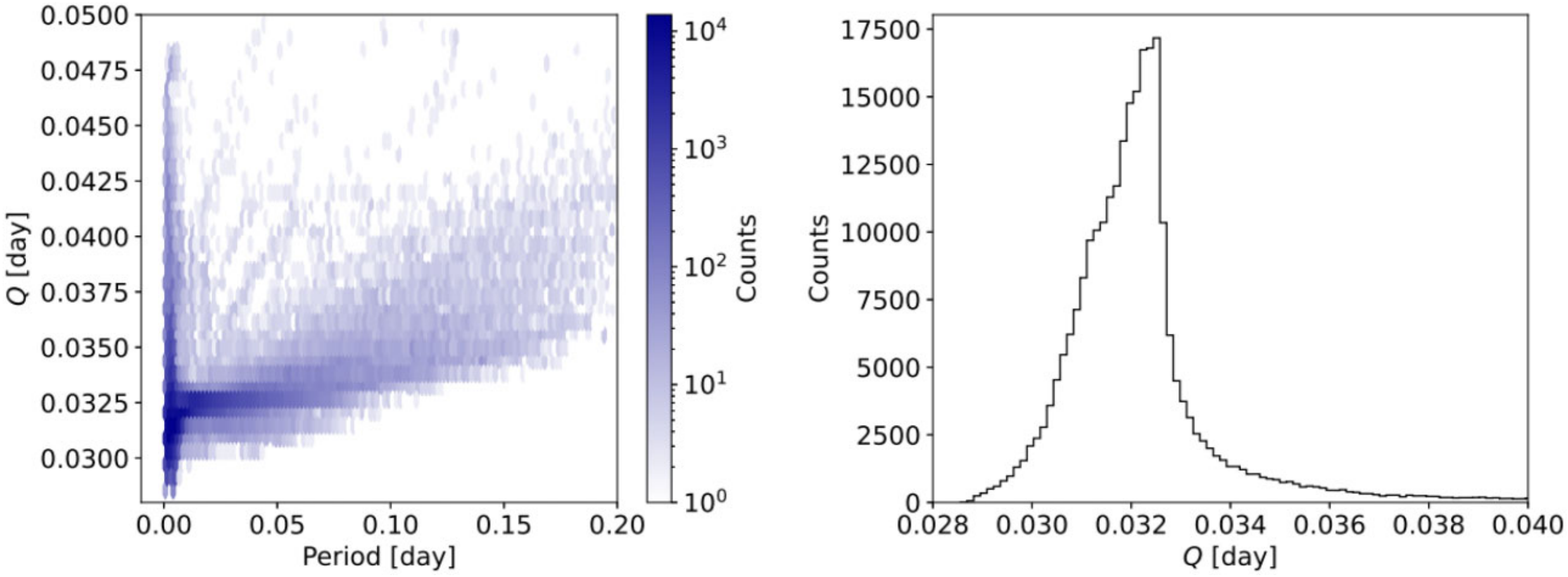

**Figure 15.** The distributions of the pulsation constant $Q$ for all the calculated radial fundamental modes.

M. Breger 1979; M. Breger et al. 1999). For the radial fundamental modes of $\delta$ Scuti stars, $Q = 0.033$ d (M. Breger & M. Montgomery 2000). Additionally, M. Breger et al. (1999) reported that, for $M = 1.80$–$1.90\,\mathrm{M}_{\odot}$ in the temperature range $T_{\mathrm{eff}} = 7400$–$7600$, the pulsation constant $Q$ is about 0.0326 with a relative accuracy of about 20 per cent for the radial fundamental modes. Here, for all the calculated radial fundamental modes, $Q$ spans a broad range from 0.029 to 0.050, but most concentrate around 0.0324, which is close to that of $\delta$ Scuti stars, as shown in Fig. 15.

For p-mode oscillations, there are relations:

$$\nu_{nl} \simeq (n + \frac{l}{2} + \epsilon)\Delta\nu, \tag{7}$$
$$\Delta\nu \sim \bar{\rho}^{1/2} \sim M^{1/2} R^{-3/2},$$

(refer to e.g. M. Tassoul 1980; D. O. Gough 1990; H. Kjeldsen & T. R. Bedding 1995; C. Aerts et al. 2010), where $\nu_{nl}$ represents the oscillation frequency, $n$ stands for the radial order of oscillation, and $l$ corresponds to the degree, while $\Delta\nu$ denotes the large frequency separation, $\epsilon$ refers to the phase constant, and $\bar{\rho}$ signifies the mean stellar density. For radial modes (i.e. $l = 0$), the period $P_{n,l=0}$ can be expressed as

$$\log P_{n,l=0} \sim \frac{3}{2}\log R - \frac{1}{2}\log M - \log(n + \epsilon), \tag{8}$$
$$\log P_{n,l=0} \sim -\frac{3}{4}\log g + \frac{1}{4}\log M - \log(n + \epsilon).$$

The above relations imply that the period $\log P$ varies with stellar radius $\log R$ or surface gravity $\log g$ when the radial order $n$ is known and ignores the effects of stellar mass.

Fig. 16 illustrates that the periods of the radial oscillations, $\log P$, exhibit nearly linear variations with $\log R$ and $\log g$. These relationships are depicted in the right ($\log R$) and left ($\log g$) panels. The upper panels of Fig. 16 display all the calculated modes, while the bottom panels show all the unstable modes. We fit these models with the following relationships:

$$\log P_n = a \cdot \log R + b, \tag{9}$$
$$\log P_n = a \cdot \log g + b.$$

The fitting coefficients are listed in Table 2. The dashed lines in Fig. 16 show the fitting results.

Additionally, we analyse these models within the region of $4.3 \le \log T_{\mathrm{eff}} \le 5.0$ and $3.5 \le \log g \le 5.5$ at the HR diagram, i.e. the area of BLAPs, called as 'small box' in follows. Similarly, we analyse these unstable modes and list the corresponding fitting results in Table 2. They are named Case I, II, III, and IV. Case I – for all the calculated models with oscillation analysis; Case II–Case I with the constraints of the small box; Case III – for all the models that can excite unstable radial modes; Case IV–Case III with the constraints of the small box.

Table 2 shows that while uncertainties in fitting coefficients increase from Case I to Case IV due to decreasing sample sizes, the standard deviation of the fits decreases. The fitting coefficients regularly vary (increase or decrease monotonously) with the radial order $n$ in both Cases I and II. However, the above regularity is destroyed for the unstable modes in Cases III and IV. This is because the distributions of the models that can excite unstable radial modes vary with their radial order $n$, as shown in Figs 4, 9, 10, and 13. Table 2 further demonstrates the high precision of the fitted period–$\log g$ and period–$\log R$ relations, and the fitting results regularly follow the theoretical predictions, i.e. equation (8).

Fig. 17 reveals that unstable dipole modes exhibit a wider period distribution than radial modes, populating nearly the entire parameter space. Similar to radial modes, both their minimum and maximum periods decrease with increasing $\log g$. Their radial order $|n_{pg}|$ also decrease with increasing $\log g$.

For the $\log P - \log g$ relation of BLAPs, P. Pietrukowicz et al. (see figs 12 and 14 of 2024) firstly made a linear fit ($\log g = a \cdot \log P + b$) for the observations and obtained $\log g = -1.14(5)\log P + 6.30(7)$ with 19 BLAPs (including four hg-BLAPs and 15 BLAPs observed by themselves with MagE spectrograph) and $\log g = -1.309(1)\log P + 6.477(1)$ with all of 24 BLAPs. For different samples, the fitting results are far apart. The slope $a$ varies from $-1.14$ to $-1.309$, and the intercept $b$ from 6.30 to 6.477 when the fitting samples include the other five BLAPs. It is worth noting that the latter fitting results seem closer to the theoretical prediction ($-4/3$) by equation (8). Based on the linear fittings and some theoretical calculations of pre-WD evolutionary models, as shown in fig. 14 of P. Pietrukowicz et al. (2024), three of them are thought of as the first overtone mode. They are TMTS-BLAP-1 (J. Lin et al. 2022), OGLE-BLAP-009 (C. W. Bradshaw et al. 2024), and SMSS-BLAP-1 (S.-W. Chang et al. 2024).

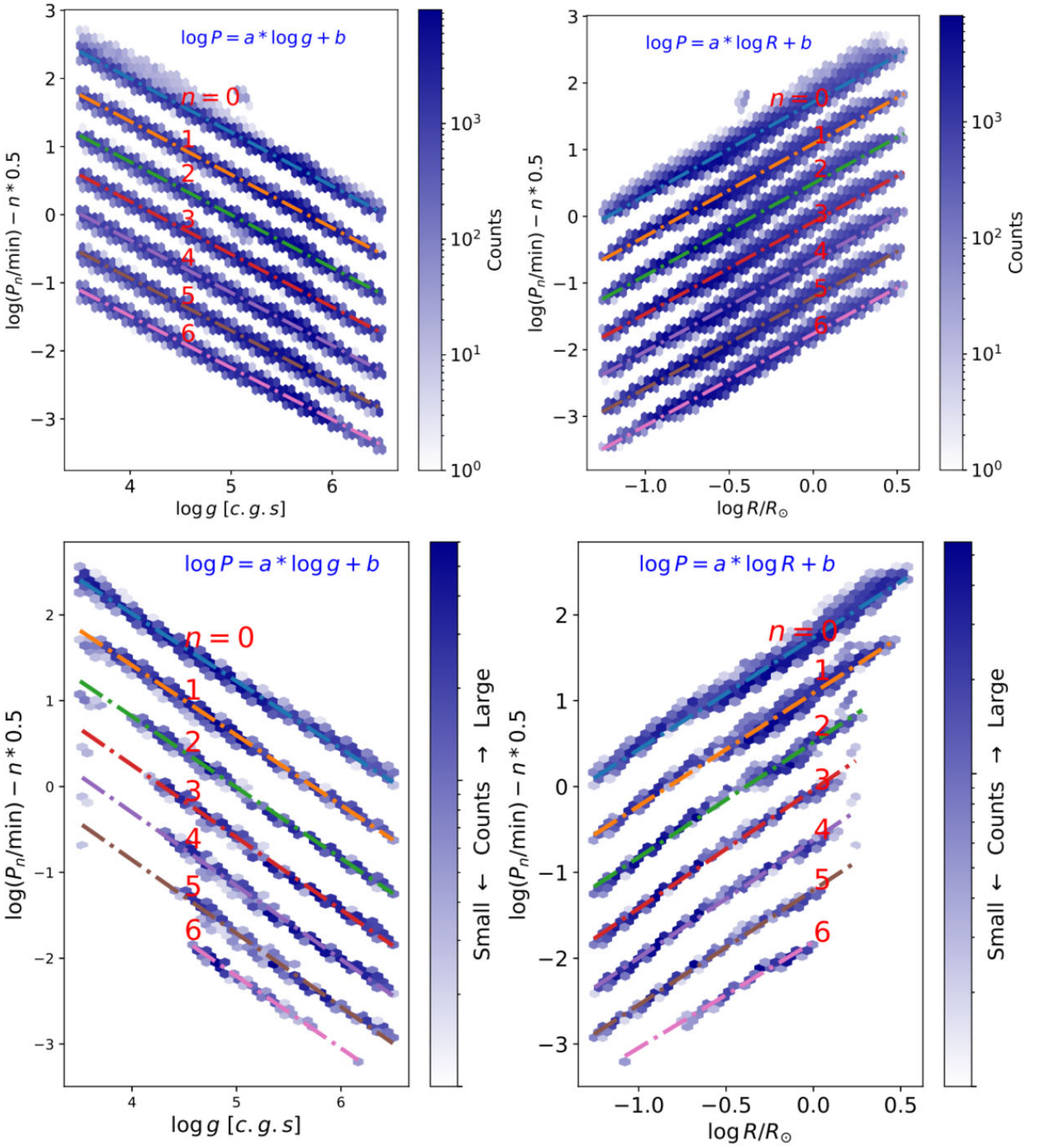

**Figure 16.** log $P$ of radial modes as a function of log $g$ (left panels) and log $R$ (right panels) for all the calculated models (upper panels) and the models whose radial modes are unstable (bottom panels). These seven series correspond to the radial fundamental mode $n = 0$, the first overtone $n = 1$, ..., and the sixth overtone $n = 6$, respectively. These dashed lines are the linear fitting results with the relation of $\log P = a \cdot \log g + b$ (left panels) and $\log P = a \cdot \log R + b$ (right panels). The fitting coefficients are listed in Table 2. Notably, in the bottom panels, the colour pattern does not denote absolute numbers for all seven series. It characterizes the relative distribution within every series. The numbers of modes with different radial orders are listed in Table 2.

Here, similar to the work of P. Pietrukowicz et al. (2024), we also fit our models (for all four cases mentioned above) with the following relations:

$$\log g = a \cdot \log P_n + b, \quad (10)$$

$$\log R = a \cdot \log P_n + b.$$

The fitting results are also listed in Table 2. The fitting coefficient precision and the corresponding standard deviation variations are similar to the fittings of equation (9).

The up-to-date observations of the BLAPs are shown in Fig. 18. The upper panel shows the unstable low-order radial modes ($n = 0, 1, 2$), i.e. the radial fundamental, first overtone, and second overtone modes. The bottom panel shows the average periods of the lowest and highest periods of the unstable dipole modes, i.e. $\frac{P_L+P_H}{2}$. Fig. 18 shows that, similar to the theoretical model prediction, log $g$ decreases with the increase of $P$, but with large dispersion, which differs from fig. 12 of P. Pietrukowicz et al. (2024). The distribution width of the observations is much broader than that of all three low-order modes, which means the BLAPs might not exhibit a single pulsation pattern; in other words, the pulsation of BLAPs might include the radial fundamental modes and other higher-order modes, such as the first and second overtone modes, if the precision and accuracy of the observations are credible. More observations and pulsation modeling are needed to test it.

**Table 2.** Fitting results for period–log $g$ and period–log $R$ relations of radial modes.

| | | | $\log P = a \cdot \log g + b$ | | | $\log g = a \cdot \log P + b$ | | | $\log P = a \cdot \log R + b$ | | | $\log R = a \cdot \log P + b$ | | |
|---|---|---|---|---|---|---|---|---|---|---|---|---|---|---|
| ID | Num. | $P$ range | $a$ | $b$ | Std. | $a$ | $b$ | Std. | $a$ | $b$ | Std. | $a$ | $b$ | Std. |
| Case I: All calculated models (including unstable and stable modes) | | | | | | | | | | | | | | |
| $f_0$ | 258759 | [0.9–503.1] | −0.7907(2) | 5.1632(8) | 0.050 | −1.2522(2) | 6.5177(3) | 0.062 | 1.3974(5) | 1.7113(3) | 0.082 | 0.6965(2) | −1.2062(2) | 0.058 |
| $f_1$ | 258257 | [0.7–209.5] | −0.7820(2) | 4.9990(8) | 0.049 | −1.2664(2) | 6.3819(2) | 0.062 | 1.3846(4) | 1.5861(3) | 0.075 | 0.7056(2) | −1.1315(2) | 0.054 |
| $f_2$ | 257949 | [0.6–167.2] | −0.7786(2) | 4.8901(8) | 0.050 | −1.2712(3) | 6.2709(2) | 0.064 | 1.3802(4) | 1.4930(3) | 0.072 | 0.7090(2) | −1.0700(2) | 0.052 |
| $f_3$ | 257700 | [0.5–140.2] | −0.7760(2) | 4.8005(8) | 0.049 | −1.2757(3) | 6.1774(2) | 0.063 | 1.3767(4) | 1.4151(2) | 0.070 | 0.7117(2) | −1.0180(2) | 0.050 |
| $f_4$ | 257447 | [0.4–119.4] | −0.7734(2) | 4.7214(8) | 0.049 | −1.2801(3) | 6.0971(2) | 0.063 | 1.3726(4) | 1.3477(2) | 0.069 | 0.7143(2) | −0.9731(2) | 0.049 |
| $f_5$ | 253919 | [0.4–103.8] | −0.7690(2) | 4.6426(8) | 0.049 | −1.2872(3) | 6.0295(2) | 0.063 | 1.3685(4) | 1.2887(2) | 0.068 | 0.7164(2) | −0.9335(2) | 0.049 |
| $f_6$ | 219426 | [0.4–91.3] | −0.7601(2) | 4.5524(9) | 0.047 | −1.3021(3) | 5.9812(2) | 0.062 | 1.3791(4) | 1.2396(3) | 0.069 | 0.7091(2) | −0.8897(2) | 0.050 |
| Case II: Models within $4.3 \leq \log T_{\text{eff}} \leq 5.0$ and $3.5 \leq \log g \leq 5.5$ on the HR diagram | | | | | | | | | | | | | | |
| $f_0$ | 114516 | [5.5–503.1] | −0.8063(3) | 5.2479(15) | 0.047 | −1.2214(4) | 6.4847(6) | 0.058 | 1.4211(9) | 1.7094(4) | 0.084 | 0.6706(4) | −1.1612(6) | 0.057 |
| $f_1$ | 114427 | [4.1–209.5] | −0.8047(2) | 5.1137(12) | 0.038 | −1.2307(4) | 6.3410(4) | 0.046 | 1.4195(9) | 1.5825(4) | 0.077 | 0.6759(4) | −1.0824(5) | 0.053 |
| $f_2$ | 114389 | [3.3–167.2] | −0.7980(3) | 4.9900(12) | 0.040 | −1.2392(4) | 6.2388(4) | 0.050 | 1.4105(8) | 1.4894(3) | 0.075 | 0.6817(4) | −1.0275(4) | 0.052 |
| $f_3$ | 114270 | [2.7–140.2] | −0.7882(3) | 4.8673(13) | 0.042 | −1.2527(4) | 6.1595(4) | 0.053 | 1.3973(8) | 1.4106(3) | 0.070 | 0.6910(4) | −0.9856(4) | 0.049 |
| $f_4$ | 114237 | [2.3–119.4] | −0.7809(3) | 4.7663(14) | 0.044 | −1.2624(4) | 6.0870(4) | 0.056 | 1.3876(7) | 1.3426(3) | 0.067 | 0.6978(4) | −0.9469(4) | 0.047 |
| $f_5$ | 114138 | [2.1–103.8] | −0.7753(3) | 4.6816(14) | 0.046 | −1.2703(5) | 6.0219(4) | 0.058 | 1.3795(7) | 1.2831(3) | 0.064 | 0.7030(4) | −0.9116(3) | 0.046 |
| $f_6$ | 102256 | [1.9–91.3] | −0.7642(3) | 4.5825(14) | 0.043 | −1.2909(5) | 5.9819(4) | 0.056 | 1.3808(8) | 1.2316(3) | 0.065 | 0.7026(4) | −0.8742(3) | 0.046 |
| Case III: Models with unstable radial modes | | | | | | | | | | | | | | |
| $f_0$ | 17342 | [1.0–363.0] | −0.7852(4) | 5.1539(20) | 0.040 | −1.2671(7) | 6.5543(11) | 0.051 | 1.3032(15) | 1.7344(7) | 0.082 | 0.7511(8) | −1.3075(13) | 0.063 |
| $f_1$ | 9147 | [0.7–162.6] | −0.8133(5) | 5.1646(26) | 0.032 | −1.2249(8) | 6.3446(11) | 0.039 | 1.3159(19) | 1.5895(9) | 0.070 | 0.7461(11) | −1.1907(14) | 0.053 |
| $f_2$ | 2830 | [0.6–118.8] | −0.8257(10) | 5.1190(50) | 0.037 | −1.2065(14) | 6.1961(14) | 0.045 | 1.3405(25) | 1.5159(17) | 0.060 | 0.7385(14) | −1.1247(14) | 0.044 |
| $f_3$ | 2013 | [0.5–91.1] | −0.8425(13) | 5.1207(71) | 0.038 | −1.1813(18) | 6.0749(14) | 0.045 | 1.3839(24) | 1.4640(18) | 0.042 | 0.7184(12) | −1.0555(10) | 0.030 |
| $f_4$ | 1686 | [0.4–76.0] | −0.8547(18) | 5.1167(97) | 0.045 | −1.1616(24) | 5.9827(17) | 0.052 | 1.3728(27) | 1.3790(21) | 0.042 | 0.7238(14) | −1.0024(10) | 0.031 |
| $f_5$ | 1214 | [0.4–65.1] | −0.8549(29) | 5.0643(155) | 0.057 | −1.1541(39) | 5.9167(26) | 0.067 | 1.3380(31) | 1.2945(23) | 0.040 | 0.7426(17) | −0.9654(12) | 0.030 |
| $f_6$ | 390 | [0.6–14.4] | −0.8399(44) | 4.9912(220) | 0.027 | −1.1781(62) | 5.9326(53) | 0.032 | 1.2336(78) | 1.1895(29) | 0.032 | 0.7984(50) | −0.9543(43) | 0.026 |
| Case IV: Unstable radial modes within Case II parameter space | | | | | | | | | | | | | | |
| $f_0$ | 14036 | [6.1–363.0] | −0.8249(5) | 5.3258(23) | 0.031 | −1.2059(7) | 6.4457(12) | 0.038 | 1.4216(20) | 1.7273(6) | 0.071 | 0.6842(10) | −1.1838(16) | 0.049 |
| $f_1$ | 7174 | [5.0–162.6] | −0.8286(9) | 5.2365(41) | 0.030 | −1.1971(13) | 6.3057(19) | 0.036 | 1.4398(32) | 1.5885(8) | 0.061 | 0.6707(15) | −1.0688(22) | 0.042 |
| $f_2$ | 1280 | [8.0–68.5] | −0.8025(30) | 5.0329(135) | 0.027 | −1.2248(45) | 6.2426(63) | 0.033 | 1.2932(103) | 1.4910(18) | 0.057 | 0.7148(57) | −1.0726(79) | 0.042 |
| $f_3$ | 697 | [5.1–43.5] | −0.8168(41) | 5.0254(191) | 0.027 | −1.2033(60) | 6.1268(74) | 0.033 | 1.2157(66) | 1.4231(16) | 0.029 | 0.8063(43) | −1.1508(54) | 0.024 |
| $f_4$ | 551 | [3.6–31.7] | −0.8459(43) | 5.1069(204) | 0.026 | −1.1656(59) | 6.0191(66) | 0.030 | 1.1777(63) | 1.3300(17) | 0.027 | 0.8360(45) | −1.1149(50) | 0.023 |
| $f_5$ | 443 | [2.7–20.8] | −0.8438(47) | 5.0535(227) | 0.025 | −1.1690(65) | 5.9730(66) | 0.030 | 1.1880(78) | 1.2505(22) | 0.029 | 0.8261(54) | −1.0371(55) | 0.025 |
| $f_6$ | 361 | [2.5–14.4] | −0.8442(60) | 5.0119(296) | 0.026 | −1.1633(83) | 5.9188(73) | 0.030 | 1.1700(93) | 1.1756(29) | 0.029 | 0.8356(67) | −0.9883(59) | 0.024 |

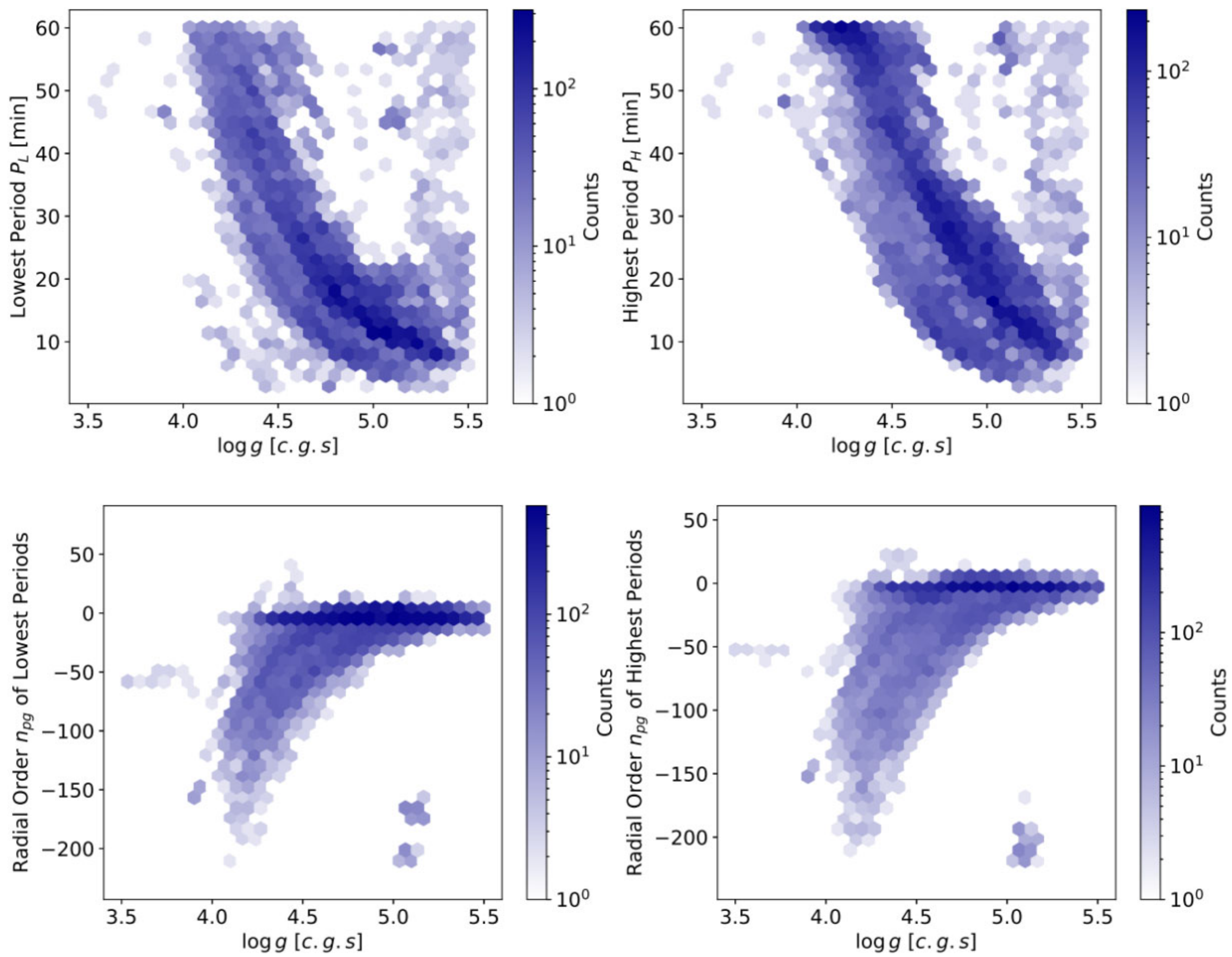

**Figure 17.** Periods $P$ and radial order $n_{pg}$ of the unstable dipole modes as a function of log $g$ for the models within the parameter space of log $T_{\rm eff}$ >= 4.3 and log $g$ <= 5.5 at the HR diagram.

### 3.4 Comparisons between with and without heavy element gravity settling and radiative levitation

The effect of gravity settling of heavy elements is that the heavier elements accumulate towards the centre of the star and the lighter elements rise to stellar surface. Finally, the atmosphere is dominated by light element(s). For the extreme horizontal branch stars, gravity settling will lead that the outer envelope becomes a pure hydrogen atmosphere quickly, and all of the heavier elements, including from helium to nickel, disappear from stellar surface, if the star has hydrogen. On the contrary, the effect of the radiative levitation will partly push the heavy elements to stellar envelope. Therefore, the inner structure and the evolutionary tracks of stars will be slightly changed.

In order to analyse the influence of the gravity settling and radiative levitation for our results, we build a set of models with the initial parameter spaces of $Z_{\rm init} = 0.04$, $M_{\rm He\text{-}core,init} = 0.7$ M$_\odot$, $M_{\rm env,init} = (0.005, 0.01, 0.02, 0.05, 0.1)$ M$_\odot$, $X_{\rm init} = (0.2, 0.4, 0.6, 0.8)$, and with the GS98 metal compositions and OP opacity table, which are illustrated in Fig. 19. The other model settings are same with the previous analysis. In these models, all elements which are heavier than hydrogen, i.e. from the helium to nickel, are considered for calculating gravity settling and radiative levitation. In addition, we select a set of reference models with the same parameter spaces from the previous model sets, which are displayed in Fig. 20.

Fig. 19 shows that the radial fundamental modes ($f_0$) are unstable for almost all core-helium-burning models. While merely a few hotter late-evolved models, i.e. shell-helium-burning models, can excite the radial fundamental modes. For the fist radial overtone ($f_1$) and other high-order overtone ($f_2$–$f_6$), the areas where the model can excite pulsation instability on the HR diagram shrink towards the hot models along with the increase of the mode orders. Notably, these hottest models can excite all analysed radial modes, i.e. $f_0$–$f_6$.

Compared to the reference models, models with gravity settling and radiative levitation more likely excite pulsations. Especially for high-order modes ($f_3$–$f_6$), all reference models can not excite any pulsations, while these models with gravity settling and radiative levitation can partly excite. This means that gravity settling and radiative levitation is beneficial to exciting pulsation in such kind stars and to explaining the pulsation observations. This is because that the gravity settling and radiative levitation leads that the iron and nickel are enriched around $\log T/K \sim 5.0 - 5.5$, which enhances the opacity. This is the energy source of the $\kappa-$mechanism. Simultaneously, it brings another trouble for explaining the composition of BLAP outer envelope. The gravity settling and radiative levitation lead that the outer envelope is consist of hydrogen (dominant) and iron and nickel. At least, the current observations show that part of BLAPs have rich helium atmosphere.

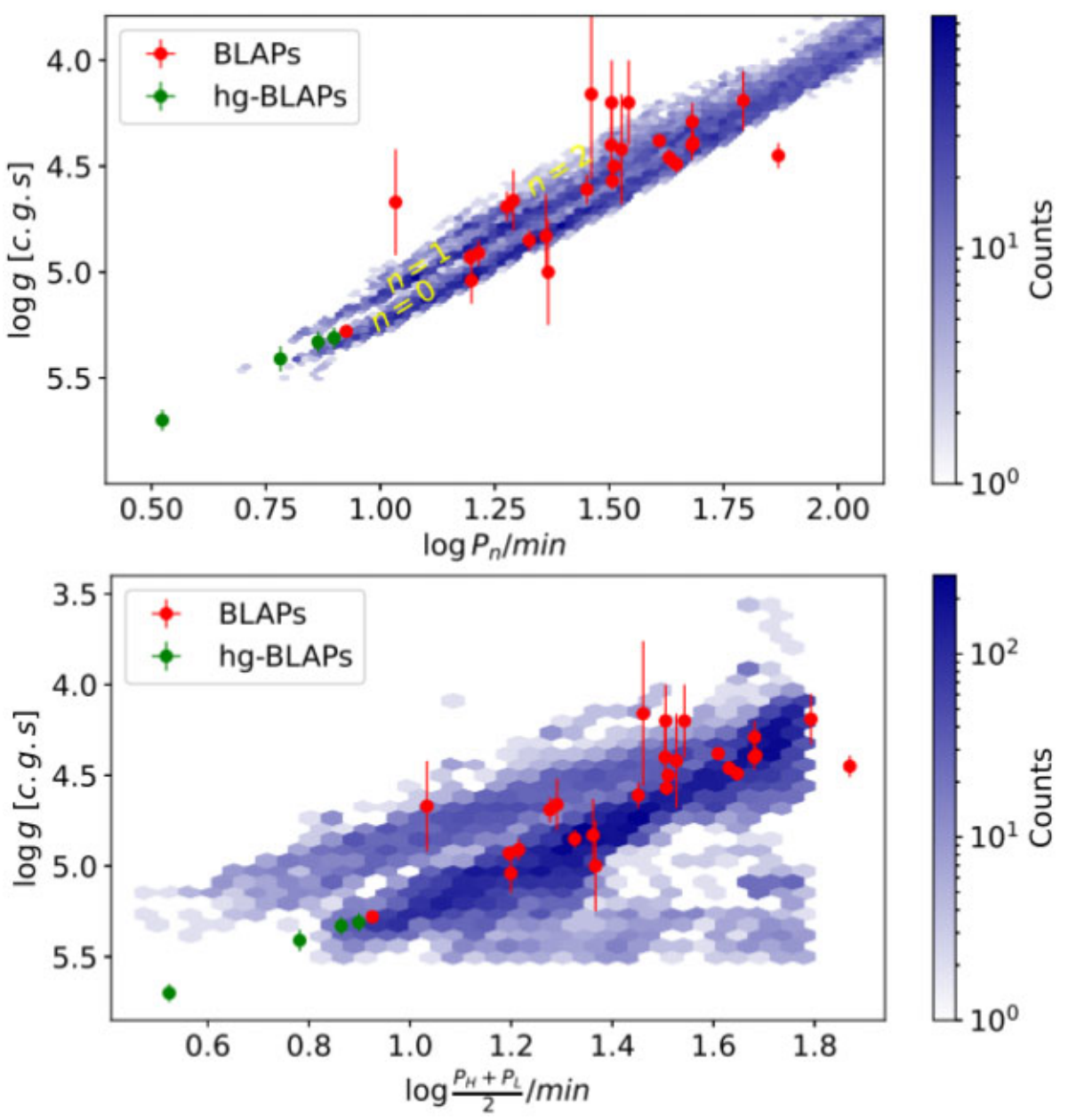

**Figure 18.** Upper panel: $\log g$ as a function of period $P_n$ for low order unstable radial modes ($n = 0,\ 1,$ and 2); Bottom panel: $\log g$ as a function of the average period ($\frac{P_H+P_L}{2}$) of the highest and lowest periods of the unstable dipole modes. Similar to the dipole modes, the radial modes used here are selected with $\log T_{\rm eff} >= 4.3$ and $\log g <= 5.5$ at the HR diagram not all of unstable modes. Red and green circles are the observations of the normal BLAPs and hg-BLAPs, respectively, which are listed in Table 1.

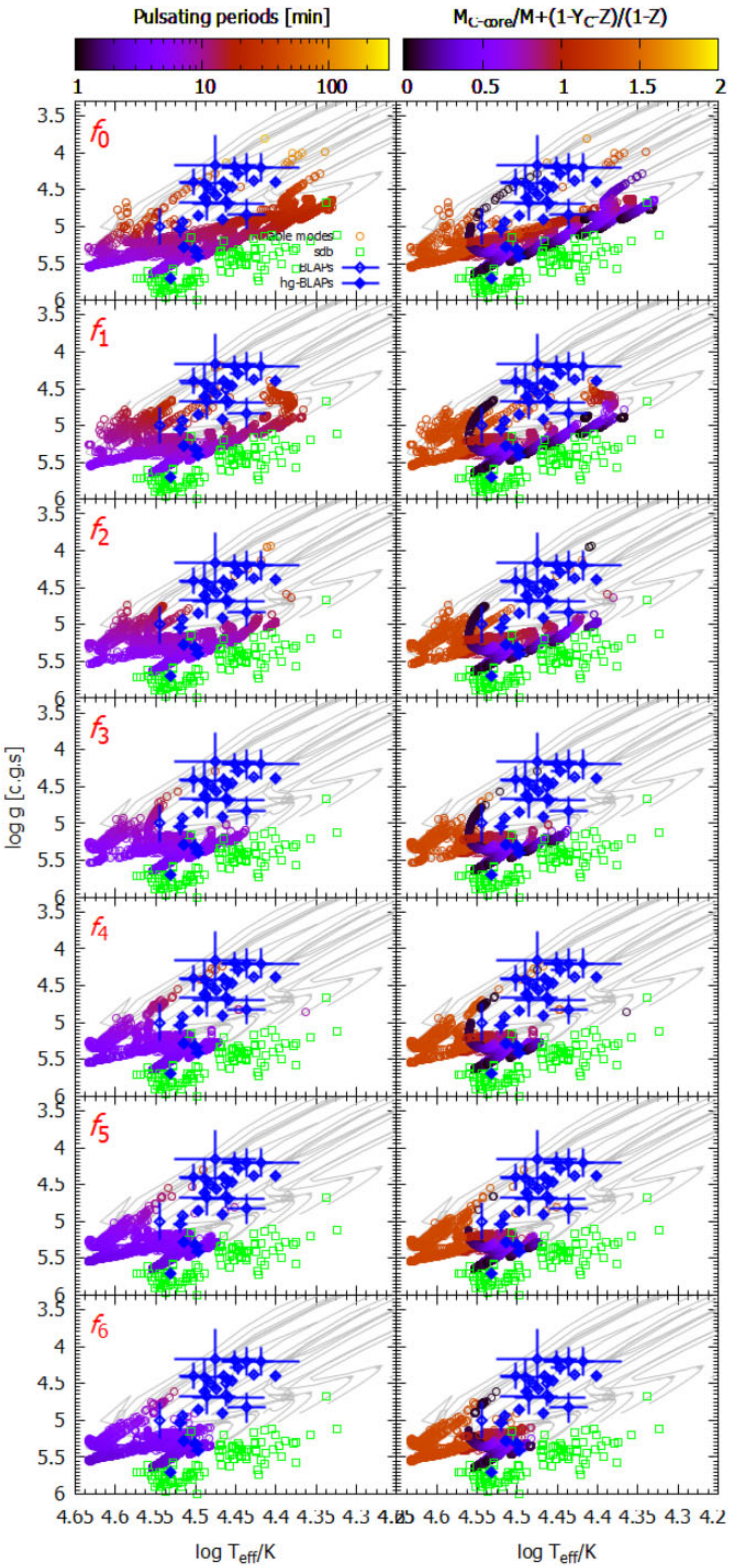

**Figure 19.** Similar to Figs 4, 9, and 10, but for the models with gravity settling and radiative levitation of heavy elements. From the top the the bottom panels, they are $f_0$, $f_1$, ..., and $f_6$, respectively. These theoretical models adopt the GS98 metal compositions and OP opacity table, i.e. OP-GS98. The other initial parameter spaces are $Z_{\rm init} = 0.04$, $M_{\rm He\text{-}core,init} = 0.7\ {\rm M}_\odot$, $M_{\rm env,init} = (0.005, 0.01, 0.02, 0.05, 0.1)\ {\rm M}_\odot$, and $X_{\rm init} = (0.2, 0.4, 0.6, 0.8)$.

## 4 CONCLUSIONS AND DISCUSSIONS

To investigate BLAP properties, we computed a theoretical model grid with the GS98 metal composition (N. Grevesse & A. J. Sauval 1998) and the OP opacity table (M. J. Seaton 2005), i.e. OP-GS98. First, we generated a compact helium core ($M_{\rm He\text{-}core,init} = 0.35$–$1.30\ {\rm M}_\odot$) with a given metal mass fraction ($Z_{\rm init} = 0.01, 0.02,$ and $0.04$). Then, we created the initial model by accreting a thin envelope with a specified hydrogen mass fraction ($X_{\rm init} = 0.01$–$0.95$) and envelope thickness ($M_{\rm env,init} = 0.001$–$0.1{\rm M}_\odot$). Subsequently, we calculated the non- and adiabatic pulsation information of these models within the region of $3.8 \leq \log T_{\rm eff} \leq 5.0$ and $3.5 \leq \log g \leq 6.5$ on the HR diagram for both low-order radial and dipole modes. For radial modes, we calculated the fundamental modes ($f_0$), first overtone ($f_1$), up to sixth overtone ($f_6$). Regarding dipole modes, we calculated all modes ranging from 3 to 60 min. Our theoretical models do not account for element diffusion and radiative levitation. Based on the theoretical model grid, we examined their pulsation instability, period changes, period–$\log g$ relation, period–$\log R$ relation, etc., and drew the following conclusions:

(i) Unstable radial fundamental ($f_0$) and first overtone ($f_1$) modes successfully replicate BLAP observations on the HR diagram (excluding high-gravity hg-BLAP-01). Within the BLAP parameter space, $f_0$ modes exhibit periods of 6–300 min, consistent with most normal BLAPs. The unstable second overtone ($f_2$) modes partially match the BLAPs observations, while higher-order modes (i.e. $f_3$–$f_6$) cannot be excited within the BLAPs region.

(ii) A richer metal abundance favors pulsation excitations, while the helium and hydrogen composition and the outer envelope thickness do not impact pulsation excitations. Relative to the metal compositions GN93 (N. Grevesse & A. Noels 1993) and GS98 (N. Grevesse & A. J. Sauval 1998), A09 (M. Asplund et al. 2009) better explains BLAPs observations. Regarding opacity tables, OP (M. J. Seaton 2005) outperforms OPAL (C. A. Iglesias & F. J. Rogers 1996). Among various combinations, OP-A09 emerges as the most favourable. For GS98 (OPAL + GS98) and GN93 (OPAL + GN93) combinations, the second overtone ($f_2$) and higher-order radial modes ($f_3$–$f_6$) fail to be excited in the BLAPs region, while the other combinations can partially excite the second overtone modes ($f_2$).

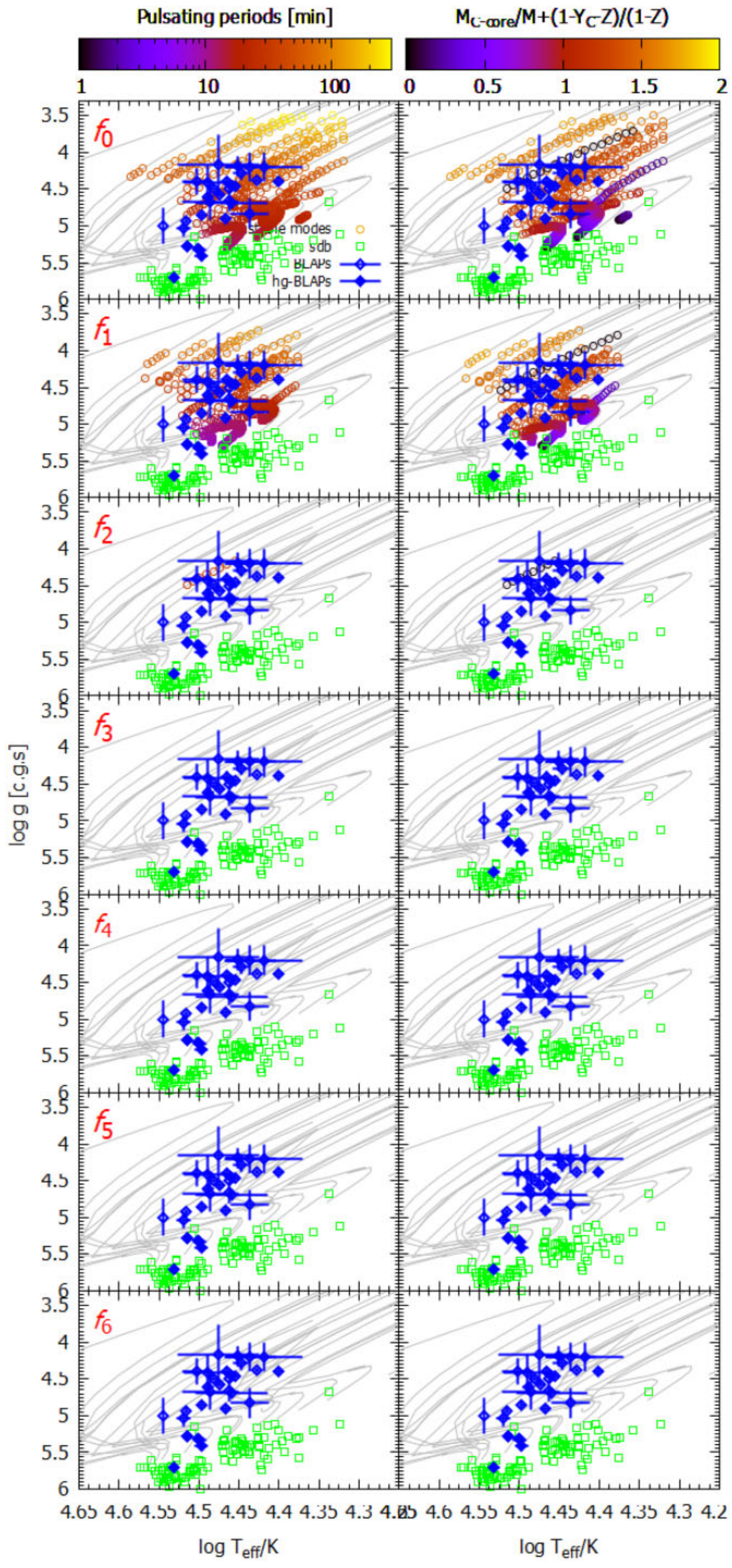

**Figure 20.** Same with Fig. 19, but without gravity settling and radiative levitation.

(iii) Concerning dipole modes, models capable of exciting unstable modes distribute into a vast region on the HR diagram. The average periods of the lowest and highest unstable modes regularly vary with surface gravity. The corresponding radial order $n_{pg}$ is predominantly between −10 and 2, with a concentration around −2. However, using the dipole modes to explain BLAPs observations proves challenging.

(iv) In theory, the period of radial modes exhibits a linear dependence on stellar surface gravity (log $g$) and radius (log $R$). However, observations reveal that BLAPs occupy a broader parameter band. This suggests that their pulsation may not be governed solely by a single mode (e.g. the radial fundamental mode); higher-order modes, such as first and second overtones, are also possible. To identify the dominant pulsation modes, we are modelling BLAPs light curves and radial velocity curves using radial stellar pulsation (RSP) theory (e.g. R. Smolec & P. Moskalik 2008; R. Smolec 2016).

(v) Although the pulsation constant $Q$ spans a wide range (0.029–0.050 d), values predominantly cluster near 0.0324 days. This closely aligns with the characteristic $Q$-value of 0.0326 d for $\delta$ Scut stars (M. Breger et al. 1999).

In summary, in this work, whether using $Z = 0.01$, 0.02, or 0.04, the position of BLAPs at the HR diagram can excite pulsation instability for the radial modes, especially for the radial fundamental modes. A richer metallicity is more beneficial for explaining the observations. Whether the positions at the HR diagram, periods, or period changes of the unstable modes, the theoretical models can perfectly describe the BLAPs observations, besides a few high-gravity hg-BLAPs. The theoretical models form a narrow band in the log $P$– log $g$ plot, while the observations exhibit a wider spread. In the future, we plan to incorporate Gaia observations (Gaia Collaboration 2016, 2023) to determine stellar radii and validate theoretical models using the period-log $R$ relation. Unfortunately, only a few targets currently have *Gaia* data; we eagerly await the next available data release.

## ACKNOWLEDGEMENTS

This work is co-supported by the National Natural Science Foundation of China (grant no. 12288102), the National Key R&D Program of China (grant no. 2021YFA1600400/2021YFA1600402), and the B-type Strategic Priority Programme of the Chinese Academy of Sciences (grant no. XDB1160202). The authors also gratefully acknowledge the supports of NSFC of China (grant nos. 12133011 and 12273104), Yunnan Fundamental Research Projects (grant no. 202401AS070045), Youth Innovation Promotion Association of Chinese Academy of Sciences, Ten Thousand Talents Program of Yunnan for Top-notch Young Talents, the International Centre of Supernovae, Yunnan Key Laboratory (no. 202302AN360001), and China Manned Space Program (grant no. CMS-CSST-2025-A14/A01). The authors express their sincere thanks to OGLE and ZFT team for these observations making this work possible and also gratefully acknowledge the computing time granted by the Yunnan Observatories, and provided on the facilities at the Yunnan Observatories Supercomputing Platform and the 'PHOENIX Supercomputing Platform' jointly operated by the Binary Population Synthesis Group and The Stellar Astrophysics Group at Yunnan Observatories, Chinese Academy of Sciences.

## DATA AVAILABILITY

Codes and data used in this work are available upon reasonable request to the author, Tao WU (email: wutao@ynao.ac.cn).

This paper has been typeset from a TEX/LATEX file prepared by the author.